\documentclass[12pt]{article}

\pdfoutput=1
\usepackage[latin9]{inputenc}
\usepackage[top=80pt,bottom=85pt,left=67pt,right=67pt]{geometry}
\usepackage{amssymb,amsmath}
\usepackage{xypic,subfigure}
\usepackage{graphicx,color}
\usepackage{float}
\usepackage{cite}
\usepackage[debug,pageanchor=false]{hyperref}
\definecolor{link}{rgb}{.8,.15,.1}
\hypersetup{colorlinks=true,linkcolor=link,citecolor=link,urlcolor=link,linktocpage}

\makeatletter
\@addtoreset{equation}{section}
\makeatother

\renewcommand{\theequation}{\thesection.\arabic{equation}}

\newcommand{\beq}{\begin{equation}}
\newcommand{\eeq}{\end{equation}}
\newcommand{\bea}{\begin{eqnarray}}
\newcommand{\eea}{\end{eqnarray}}

\newcommand{\eq}{\begin{equation}}
\newcommand{\feq}{\end{equation}}
\newcommand{\eqn}{\begin{eqnarray}}
\newcommand{\feqn}{\end{eqnarray}}

\newcommand{\ma}[1]{\mbox{$\mathcal{#1}$}}

\newcommand{\mrm}[1]{\mbox{$\mathrm{#1}$}}

\begin{document}
\begin{titlepage}

\begin{center}

\vskip .5in 
\noindent

{\Large \bf{Searching for surface defect CFTs within AdS$_3$}}

\bigskip\medskip

Federico Faedo$^a$\footnote{federico.faedo@unimi.it}, Yolanda Lozano$^b$\footnote{ylozano@uniovi.es},  Nicol\`o Petri$^b$\footnote{petrinicolo@uniovi.es} \\
\bigskip\medskip
{\small 

a: Dipartimento di Fisica, Universit\`a di Milano, and \\
INFN, Sezione di Milano, \\
Via Celoria 16, I-20133 Milano, Italy.}

\bigskip\medskip
{\small 

b: Department of Physics, University of Oviedo,
Avda. Federico Garcia Lorca s/n, 33007 Oviedo, Spain.}

\vskip 1cm 

     	{\bf Abstract }
     	\end{center}
     	\noindent

We study $\mathrm{AdS}_3\times S^3/\mathbb{Z}_k\times {\tilde S}^3/\mathbb{Z}_{k'}$ solutions to M-theory preserving $\mathcal{N}=(0,4)$ supersymmetries, arising as near-horizon limits of M2-M5 brane  intersections ending on M5'-branes, with both types of five-branes placed on A-type singularities. Solutions in this class
asymptote locally to $\mathrm{AdS}_7/\mathbb{Z}_k\times {\tilde S}^3/\mathbb{Z}_{k'}$, and can thus be interpreted as holographic duals to surface defect CFTs within the $\mathcal{N}=(1,0)$ 6d CFT dual to this solution. Upon reduction to Type IIA, we obtain a new class of solutions of the form $\mathrm{AdS}_3\times S^3/\mathbb{Z}_k\times  S^2 \times \Sigma_2$ preserving (0,4) supersymmetries. We construct explicit 2d quiver CFTs dual to these solutions, describing D2-D4 surface defects embedded within the 6d (1,0) quiver CFT dual to the $\mathrm{AdS}_7/\mathbb{Z}_k$  solution to massless IIA. Finally, in the massive case, we show that the recently constructed
$\mathrm{AdS}_3\times S^2\times \mathrm{CY}_2$ solutions with $\mathcal{N}=(0,4)$ supersymmetries gain a defect interpretation as surface CFTs originating from D2-NS5-D6 defects embedded within the 5d CFT dual to the Brandhuber-Oz $\mathrm{AdS}_6$ background. 

\noindent

\vfill
\eject

\end{titlepage}

\setcounter{footnote}{0}

\tableofcontents

\setcounter{footnote}{0}
\renewcommand{\theequation}{{\rm\thesection.\arabic{equation}}}

\section{Introduction}

The crucial ingredient of the AdS/CFT regime of string theory is that it provides a concrete set-up where it is possible to handle (some) non-perturbative effects featuring the gravitational interaction at its quantum phase \cite{Maldacena:1997re,Witten:1998qj}. In particular, the possibility of working out quantitative results of the physics of branes allowed (and still allows) to spread new light on the most interesting and mysterious features of quantum gravity, like, for example, the existence of non-Lagrangian phases for quantum fields.
Despite the huge amount of  new ideas, proposals and results that revolve around holography, issues such as the holographic interpretation of lower-dimensional AdS backgrounds are still in need of a
deeper understanding. 

A very interesting approach to the study of AdS backgrounds in lower dimensions is to resolve their  
dual CFTs within higher-dimensional field theories. In string theory this idea gains a precise realisation when the AdS geometries are part of higher-dimensional solutions with non-compact internal manifolds. When that happens one can use that the number of dynamical degrees of freedom of a holographic CFT is proportional to the coupling constant of the corresponding AdS solution, which is in turn related to the volume of the internal manifold \cite{Brown:1986nw}. From this two important lessons can be extracted. The first is that the non-compactness of the internal manifold can be considered as signalling the presence of an underlying higher-dimensional field theory. The second is that the partial breaking of the Lorentz (and, in case, conformal) symmetries of the spacetime where the higher-dimensional field theory lives can be considered as entirely due to the presence of the AdS geometry.

Defect conformal field theories constitute a perfect framework for the implementation of these ideas \cite{Cardy:1984bb,Cardy:1991tv,McAvity:1993ue}. In this context some of the conformal isometries of a higher-dimensional CFT are broken by a deformation driven by a position-dependent coupling, implying non-vanishing 1-point functions and non-trivial displacement operator (the energy-momentum tensor is not preserved). To date, many examples of defect CFTs have been discussed in the string theory literature. For a non-exhaustive list of references see \cite{Karch:2000gx,DeWolfe:2001pq,Bachas:2001vj,Erdmenger:2002ex,Constable:2002xt,Aharony:2003qf,Bak:2003jk,Clark:2004sb,Kapustin:2005py,Clark:2005te,DHoker:2006qeo,DHoker:2006vfr,Buchbinder:2007ar,DHoker:2007zhm,DHoker:2007hhe,Lunin:2007ab,Gaiotto:2008sa,Gaiotto:2008sd,Aharony:2011yc,Gutperle:2012hy,Jensen:2013lxa,Estes:2014hka,deLeeuw:2015hxa,Billo:2016cpy,Dibitetto:2017klx,DelZotto:2018tcj,Lozano:2019ywa}. The defect CFTs usually come about when a brane intersection ends on a bound state which is known to be described by an AdS vacuum in the near-horizon limit. The intersection breaks some of the isometries of the vacuum, producing a lower-dimensional AdS solution described by a non-trivial warping between AdS and the internal manifold. The defect describes then the boundary conditions associated to the intersection between the defect branes and the original bound state.

A very useful approach to the study of these systems comes from their description in lower-dimensional supergravities. A simple reason for this is that the parametrisation of an AdS string solution often hides the presence of higher-dimensional AdS vacua, that may describe the background in some particular limit. Instead, in lower dimensions one can directly search for solutions in which the defect interpretation is manifest. More concretely, given an $\mathrm{AdS}_d$ vacuum associated to a particular brane system, one can consider $d$-dimensional Janus-type backgrounds
\begin{equation}
 ds_d^2=e^{2U(\mu)}\,ds^2_{{\scriptsize \mrm{AdS}_{p+2}}}+e^{2W(\mu)}\,ds^2_{d-p-3}+e^{2V(\mu)}\,d\mu^2\,,
 \label{slicing} 
\end{equation} 
with non-compact $\ma {M}_{d-p-3}\times I_\mu$ transverse space,
 admitting an asymptotic region locally described by the $\mrm{AdS}_d$ vacuum. These backgrounds can then be consistently uplifted to 10 or 11 dimensions,  producing warped geometries of the type $\mrm{AdS}_{p+2}\times \ma {M}_{d-p-3}\times I_\mu\times \Sigma_{D-d}$, with $\Sigma_{D-d}$ the internal manifold of the truncation. Holographically, this is the supergravity picture of a defect $(p+1)$-dimensional CFT realised within a higher $(d-1)$-dimensional CFT.
 
Following this philosophy, in this paper we will be concerned with AdS$_3$, and in a lesser degree AdS$_2$, solutions with 4 supercharges, arising as near-horizons of brane intersections in M-theory and massive IIA string theory, to which we will propose a holographic interpretation in terms of defect conformal field theories.

Due to the high dimensionality of the associated internal manifolds, a complete scanning and classification of $\mathrm{AdS}_3$ and $\mathrm{AdS}_2$ backgrounds is still missing (for a non-exhaustive list of references see \cite{Argurio:2000tg,Kim:2005ez,Gauntlett:2006ns,Gauntlett:2006af,DHoker:2007mci,Donos:2008hd,Corbino:2017tfl,Corbino:2018fwb,Dibitetto:2017tve,Kelekci:2016uqv,Dibitetto:2017klx,Couzens:2017way,Eberhardt:2017uup,Couzens:2017nnr,Gauntlett:2018dpc,Dibitetto:2018gbk,Dibitetto:2018ftj,Dibitetto:2018iar,Dibitetto:2018gtk,Couzens:2018wnk,Gauntlett:2019roi,Macpherson:2018mif,Hong:2019wyi,Lozano:2019jza,Couzens:2019iog,Legramandi:2019xqd,Corbino:2020lzq,Lozano:2020bxo,Lozano:2019zvg,Cvetic:2000cj,DHoker:2008lup,DHoker:2008rje,Dibitetto:2019nyz}). Moreover, many are the examples of already known solutions in need of a clearer understanding of the physics of the non-perturbative objects that underlie them. In this paper we will focus our study on AdS$_3$ solutions with $\ma N=(0,4)$  supersymmetry\footnote{See section \ref{line-defects} for a brief account on AdS$_2$ solutions with 4 supercharges.}. 
These solutions have received renewed interest recently, having been studied in a series of papers  \cite{Couzens:2017way,Lozano:2019emq,Lozano:2019jza,Lozano:2019ywa,Lozano:2019zvg,Lozano:2020bxo}. Their significance comes from the fact that they provide explicit holographic duals to 2d $\ma N=(0,4)$ CFTs \cite{Tong:2014yna,Kim:2015gha,Putrov:2015jpa,Hanany:2018hlz}, which in turn have been shown to play a central role in the microscopical description of 5d black holes \cite{Maldacena:1997de,Vafa:1997gr,Minasian:1999qn,Castro:2008ne,Haghighat:2015ega,Couzens:2019wls} and the study of 6d (1,0) CFTs deformed away from the conformal point  \cite{Haghighat:2013tka,Gadde:2015tra}. A precise duality between $\mathrm{AdS}_3$ solutions and 2d (0,4) quiver CFTs has been described recently in \cite{Lozano:2019ywa,Lozano:2019zvg}.

We start in section \ref{Mdefect} by taking into consideration the 11d class of  $\mathrm{AdS}_3\times S^3/\mathbb{Z}_k\times \mathrm{CY}_2\times I$ $\ma N=(0,4)$ backgrounds  recently constructed in  \cite{Lozano:2020bxo}. We focus on a subclass describing the near-horizon regime of a particular set-up of M-branes consisting on M5'-branes on which M2-M5 bound states end. For more generality both types of 5-branes are placed on ALE singularities. Besides providing the full 11d brane solution reproducing the $\mathrm{AdS}_3$ background in its near-horizon limit, we derive the right parametrisation that allows to link this 11d spacetime with a 7d domain wall described by (\ref{slicing}), found in \cite{Dibitetto:2017tve,Dibitetto:2017klx}. This 7d solution reproduces  asymptotically locally in the UV an $\mathrm{AdS}_7$ geometry, while it manifests a singular behaviour in the IR  corresponding to the locus where the defect M2-M5 branes intersect the M5'-branes.

In section \ref{IIApicture} we consider the IIA regime of this system. The M5'-branes on an A-type singularity become NS5-D6 bound states, that are intersected by D2-D4 branes coming from the reduction of the M2-M5 branes. We provide the full brane solution as well as its $\mathrm{AdS}_3$ near-horizon geometry. The $\mathrm{AdS}_3$ near-horizon solution turns out to belong to a new class of $\mrm{AdS}_3$ solutions to 10d, that we present and study in generality in appendix \ref{newAdS3}. We derive the right parametrisation that allows to link the 10d spacetime with the 7d domain wall found in \cite{Dibitetto:2017tve,Dibitetto:2017klx}. We do this by directly relating the 10d solution to the uplift of the 7d domain wall to IIA supergravity. This allows us to interpret the 10d solution as describing a surface defect CFT within the 6d (1,0) CFT dual to the AdS$_7$ solution to massless IIA supergravity  \cite{Cvetic:2000cj,Apruzzi:2013yva}. We construct the 2d $\ma N=(0,4)$ quiver CFT that explicitly describes the surface defect CFT, and discuss the agreement between the field theory and holographic central charges. 

In section \ref{massiveIIA} we consider the classification of $\ma N=(0,4)$ $\mathrm{AdS}_3\times S^2\times \mathrm{CY}_2\times I$ solutions to massive IIA supergravity constructed in \cite{Lozano:2019emq}, for $\mathrm{CY}_2=T^4$. We provide the associated full brane solution, that we interpret in terms of D2-NS5-D6 branes ending on a D4-D8 bound state. We obtain the parametrisation that relates its $\mathrm{AdS}_3$ near-horizon geometry to a 6d domain wall of the type given by \eqref{slicing}, found in \cite{Dibitetto:2018iar}. This 6d solution is asymptotically locally $\mrm{AdS}_6$. This allows us to propose, in analogy to the $\mrm{AdS}_7$ case, a dual interpretation to the $\mrm{AdS}_3$ solution as a $\ma N=(0,4)$ surface defect CFT within the 5d Sp(N) CFT \cite{Seiberg:1996bd} dual to the Brandhuber-Oz  $\mrm{AdS}_6$ background \cite{Brandhuber:1999np}. 

In section \ref{line-defects} we briefly consider the realisation of the $\mrm{AdS}_2$ solutions to massive IIA supergravity recently constructed in \cite{Lozano:2020bxo} as line defect CFTs within the 5d Sp(N) CFT. We put together previous results in the literature that allow us to provide a defect interpretation following the general line of thought taken in this paper. We find that a subclass of the solutions found in \cite{Lozano:2020bxo} can be obtained as near-horizon geometries of D0-F1-D4' bound states intersecting the Brandhuber-Oz set-up. Moreover, these solutions can be linked to a 6d domain wall of the type given by (\ref{slicing}) that is asymptotically locally $\mrm{AdS}_6$. This allows, as above, to interpret them as line defects within the 5d Sp(N) CFT.

Section~\ref{conclusions} contains our conclusions and future directions. Appendix~\ref{7dsugra} contains a summary of the M-theory origin of minimal 7d $\ma N=1$ supergravity, useful for the analysis in section 2. In appendix~\ref{newAdS3} we present an extension of the new class of $\mrm{AdS}_3$ solutions to Type IIA constructed in section 3. Appendix~\ref{summary2dCFT} contains a brief account of the main properties of 2d (0,4) quiver CFTs, of utility for the analysis in section 3.3. Finally, in appendix~\ref{6dsugra} we present a brief summary of the main features of the massive IIA truncation to Romans supergravity, on which our results in sections 4 and 5 rely.

\section{Surface defects in M-theory}\label{Mdefect}

In this section we consider a particular brane set-up in M-theory consisting on M2-M5 branes ending on M5'-branes. We consider the most general case in which the 5-branes are placed on ALE singularities,  introduced by KK and KK' monopoles. We construct the explicit supergravity solution and show that it admits a near-horizon regime described by an $\mathrm{AdS}_3\times S^3/\mathbb{Z}_k \times S^3/\mathbb{Z}_{k'}  \times \Sigma_2$ background with $\ma N=(0,4)$ supersymmetry. This geometry extends a particular subclass of the solutions recently studied in \cite{Lozano:2020bxo}. 

The main aspect to note is that the coordinates in which the near-horizon limit emerges ``hide" the presence of an underlying $\mrm{AdS}_7/\mathbb{Z}_k$ vacuum arising in the UV.  In order to show this explicitly  we link the near-horizon geometry to a 7d domain wall asymptotically locally $\mathrm{AdS}_7$. This 7d solution, first worked out in \cite{Dibitetto:2017tve,Dibitetto:2017klx}, is a Janus-like flow preserving 8 real supercharges, characterised by an $\mathrm{AdS}_3$ slicing. In 11d it is featured by a non-compact internal manifold whose asymptotic behaviour reproduces locally the $\mathrm{AdS}_7$ vacuum of M5-branes on an A-type singularity. In the ``domain wall coordinates" the near-horizon geometry of our brane set-up gains a consistent description as a flow interpolating between a local $\mathrm{AdS}_7$ geometry and a singularity. The first regime corresponds to the limit in which we are far from the M2-M5 intersection, while the second is equivalent to ``zooming in" on the region where the M2-M5 branes end on the M5'-brane, breaking the isometries of the branes that generate the vacuum.

\subsection{The brane set-up}\label{Mtheorysetup}

We start considering the supergravity picture of an M2-M5 bound state ending on orthogonal M5'-branes, with the 5-branes located at singularities defined by Kaluza-Klein monopoles with charges $Q_{\text{KK}}$ and $Q_{\text{KK}'}$. This intersection, depicted in Table \ref{Table:branesinAd7}, preserves an $\mathrm{SO}(3)\times\mathrm{SO}(3)$ bosonic symmetry and 4 real supercharges.
\begin{table}[http!]
\renewcommand{\arraystretch}{1}
\begin{center}
\scalebox{1}[1]{
\begin{tabular}{c||c c|c c c | c | c | c c c | c }
branes & $t$ & $x^1$ & $r$ & $\theta^{1}$ & $\theta^{2}$ & $\chi$ & $z$ & $\rho$ & $\varphi^1$ & $\varphi^2$ & $\phi$    \\
\hline \hline
$\mrm{KK}$' & $\times$ & $\times$ & $\times$ & $\times$ & $\times$ & $\times$ & $\times$ & $-$ & $-$ & $-$  & $\text{ISO}$ \\
$\mrm{M}5$' & $\times$ & $\times$ & $\times$ & $\times$ & $\times$ & $\times$ & $-$ & $-$ & $-$ & $-$ & $-$ \\
$\mrm{M}2$& $\times$ & $\times$ & $-$ & $-$ & $-$ & $-$ & $\times$ & $-$ & $-$ & $-$ & $-$ \\
$\mrm{M}5$ & $\times$ & $\times$ & $-$ & $-$ & $-$ & $-$ & $-$ & $\times$ & $\times$ & $\times$ & $\times$ \\
$\mrm{KK}$ & $\times$ & $\times$ &$-$ & $-$ & $-$ & $\text{ISO}$ & $\times$ & $\times$ & $\times$ & $\times$  & $\times$\\
\end{tabular}
}
\end{center}
\caption{1/8-BPS brane system underlying the intersection of M2-M5 branes ending on M5'-branes with KK monopoles. $\chi$ ($\phi$) is the Taub-NUT direction of the KK (KK') monopoles.} \label{Table:branesinAd7}
\end{table}

We consider the following 11d metric
\begin{equation}
\label{brane_metric_M2M5KKM5_branesol}
\begin{split}
d s_{11}^2&=H_{\mathrm{M}5'}^{-1/3}\,\left[H_{\mathrm{M}5}^{-1/3}\,H_{\mathrm{M}2}^{-2/3}\,ds^2_{\mathbb{R}^{1,1}}+H_{\mathrm{M}5}^{2/3}\,H_{\mathrm{M}2}^{1/3}\left(H_{\text{KK}}(dr^2+r^2d s^2_{S^2})+H_{\text{KK}}^{-1}(d\chi+Q_{\mathrm{KK}}\,\omega)^2\right) \right]\\
&+H_{\mathrm{M}5'}^{2/3}\left[H_{\mathrm{M}5}^{2/3}\,H_{\mathrm{M}2}^{-2/3}\,dz^2+H_{\mathrm{M}5}^{-1/3}\,H_{\mathrm{M}2}^{1/3}\,\left(H_{\text{KK}'}(d\rho^2+\rho^2d s^2_{\tilde{S}^2})+H_{\text{KK}'}^{-1}(d\phi+Q_{\mathrm{KK}'}\,\eta)^2\right)\right] \, ,\\
\end{split}
\end{equation}
where $\omega$ and $\eta$ are defined such that $d\omega=\text{vol}_{S^2}$ and $d\eta=\text{vol}_{\tilde{S}^2}$. We take the M2-M5 branes completely localised in the worldvolume of the M5'-branes, i.e. $H_{\mathrm{M}2}=H_{\mathrm{M}2}(r)$ and $H_{\mathrm{M}5}=H_{\mathrm{M}5}(r)$. This particular charge distribution breaks the symmetry under the interchange of the two 2-spheres. This is explicit in the 4-form flux $G_{(4)}$, 
\begin{equation}
\begin{split}
\label{G4}
 G_{(4)}&=\partial_rH_{\mathrm{M}2}^{-1}\,\text{vol}_{\mathbb{R}^{1,1}}\wedge dr\wedge dz-\partial_rH_{\mathrm{M}5} r^2\, \text{vol}_{S^2}\wedge d\chi \wedge dz\\
 &+H_{\mathrm{KK}'}\,H_{\mathrm{M}2}\,H_{\mathrm{M}5}^{-1}\,\partial_z H_{\mathrm{M}5'}\rho^2\,d\rho\wedge\text{vol}_{\tilde{S}^2}\wedge d\phi -\partial_\rho H_{\mathrm{M}5'}\rho^2\,dz \wedge\text{vol}_{\tilde{S}^2}\wedge d\phi\,.
\end{split}
\end{equation}
The equations of motion and Bianchi identities of 11d supergravity are then equivalent to two independent sets of equations: one involving the M2-M5 branes and the KK monopoles,
\begin{equation}\label{11d-defectbranesEOM}
H_{\mathrm{M}2}=H_{\mathrm{M}5}\,,\qquad \nabla^2_{\mathbb{R}^3_r}\,H_{\mathrm{M}5}=0\qquad \text{with}\qquad H_{\mathrm{KK}}=\frac{Q_{\mathrm{KK}}}{r}\,,
\end{equation}
and the other describing the dynamics of M5'-branes on the ALE singularity introduced by the KK'-monopoles,
\begin{equation}\label{11d-motherbranesEOM}
\nabla^2_{\mathbb{R}^3_\rho}\,H_{\mathrm{M}5'}+H_{\text{KK}'}\,\partial_z^2\,H_{\mathrm{M}5'}=0\qquad \text{with}\qquad H_{\mathrm{KK}'}=\frac{Q_{\mathrm{KK}'}}{\rho}\,.
\end{equation}
The second equation in \eqref{11d-defectbranesEOM} can be easily solved for
\begin{equation}\label{11d-solbranes}
  H_{\mathrm{M}5}(r)=H_{\mathrm{M}2}(r)=1+\frac{Q_{\mathrm{M}5}}{r}\,,
 \end{equation}
where we introduced the M2 and M5 charges $Q_{\mathrm{M}2}$ and $Q_{\mathrm{M}5}$, that in order to satisfy \eqref{11d-defectbranesEOM} have to be equal. One way to look at our system is then in terms of M5'-KK' branes moving on the 11d background generated by  M2-M5-KK branes. The 4d transverse manifold parametrised by the coordinates $(\rho, \varphi^1,\varphi^2, \phi)$ arises as a foliation of the Lens space $\tilde{S^3}/\mathbb{Z}_{k'}$ that is obtained by modding out the $\tilde{S^3}$ with $k^\prime=Q_{\text{KK}'}$, through the change of coordinates $\rho \rightarrow 4^{-1}\,Q_{\text{KK}'}^{-1}\,\rho^2$ \cite{Cvetic:2000cj}.

It is interesting to consider the limit $r\rightarrow 0$. This is equivalent to ``zooming in" on the locus where the M2-M5 branes intersect the M5'-branes. In this limit, the worldvolume of the M5'-branes becomes $\mathrm{AdS}_3\times S^3/\mathbb{Z}_k$, with $k=Q_{\text{KK}}$, and the full 11d string background takes the form\footnote{We redefined the Minkowski coordinates as $(t,x^1)\rightarrow 2\,Q_{\mathrm{M}5}\,Q_{\text{KK}}^{1/2}\,(t,x^1)\,.$}
\begin{equation}
\label{brane_metric_M2M5KKM5_nh}
\begin{split}
d s_{11}^2&=4\,k\,Q_{\text{M}5}\,H_{\mathrm{M}5'}^{-1/3}\,\left[ds^2_{{\scriptsize \mrm{AdS}_3}}+ds^2_{S^3/\mathbb{Z}_k} \right]+H_{\mathrm{M}5'}^{2/3}\left[dz^2+d\rho^2+\rho^2 ds^2_{\tilde{S}^3/\mathbb{Z}_{k'}}\right] \, ,\\
 G_{(4)}&=8\,k\,Q_{\text{M}5}\,\text{vol}_{{\scriptsize \mrm{AdS}_3}}\wedge dz+8\,k\,Q_{\text{M}5}\,\text{vol}_{S^3/\mathbb{Z}_k}\wedge dz\\
 &+\partial_z H_{\mathrm{M}5'}\rho^3\,d\rho\wedge \text{vol}_{\tilde{S}^3/\mathbb{Z}_{k'}}
 -\partial_\rho H_{\mathrm{M}5'}\rho^3\,dz \wedge \text{vol}_{\tilde{S}^3/\mathbb{Z}_{k'}}\,.
\end{split}
\end{equation}
Here the two orbifolded 3-spheres are locally described by the metrics
\begin{equation}\label{orbifoldS3}
 ds^2_{S^3/\mathbb{Z}_k}=\frac14\left[ \left(\frac{d\chi}{k}+\omega\right)^2+ds^2_{S^2}   \right]\,\qquad \text{and}\qquad ds^2_{\tilde S^3/\mathbb{Z}_{k'}}=\frac14\left[ \left(\frac{d\phi}{k^\prime}+\eta\right)^2+ds^2_{\tilde S^2}   \right]\,.
\end{equation}
It is important to stress the relevance of the $Q_{\text{KK}}$ monopole charge dissolved in the worldvolume of the M5'-branes in recovering the near horizon geometry, given by \eqref{brane_metric_M2M5KKM5_nh}, from the general brane solution \eqref{brane_metric_M2M5KKM5_branesol}. Besides securing that the supersymmetries of the M2-M5-M5' brane set-up are broken by a half, the presence of the KK-monopoles crucially determines the emergence of the $\mathrm{AdS}_3\times S^3/\mathbb{Z}_k$ geometry associated to the smeared M2-M5 branes.

This $\mathrm{AdS}_3$ background extends the $\ma N=(0,4)$ $\mathrm{AdS}_3\times S^3/\mathbb{Z}_k \times \mathrm{CY}_2 \times I$ backgrounds recently studied in the main body of \cite{Lozano:2020bxo} (defined by equations (3.1) and (3.2) therein)\footnote{More explicitly, we recover the subclass of solutions that are obtained uplifting the solutions referred as class I in \cite{Lozano:2019emq}. These solutions are constructed in appendix B in \cite{Lozano:2020bxo}. Within this class we recover the solutions with $\mathrm{CY}_2=\mathbb{R}^4$, or $T^4$ locally, $u^\prime=0$ and $H_2=0$. The main results in   \cite{Lozano:2020bxo} refer however to the subclass of solutions for which the M5'-branes are smeared in their transverse space.}, to the case in which the M5'-branes are completely localised in their transverse space. Taking a round ${\tilde S}^3$, i.e. $k^\prime=1$, and the M2-M5 defects smeared on the $(\rho, {\tilde S}^3)$ directions, one recovers the solutions that were the focus of  \cite{Lozano:2020bxo}, with $ds^2_{\mathrm{CY}_2}=d\rho^2+\rho^2 ds^2_{\tilde{S}^3}$. Indeed, we can recast the near-horizon solution \eqref{brane_metric_M2M5KKM5_nh} in the form of \cite{Lozano:2020bxo} by choosing
\begin{equation}
k=h_8,\qquad H_{\mathrm{M}5'}=\frac{2^6Q_{\mathrm{M}5}^3\,h_8^2}{u^2}\,h_4,\qquad z=\frac{1}{4\,Q_{\mathrm{M}5}}\,\tilde{\rho}\,, \qquad \rho=\frac{u^{1/2}}{4\,Q_{\mathrm{M}5}\,h_8^{1/2}}\,\tilde r,
\end{equation}
with $H_{\mathrm{M}5'}=H_{\mathrm{M}5'}(z)$ as a result of the smearing. In the next section we will see how the extra dependence on the $\rho$ coordinate is crucial in order to reach $\mathrm{AdS}_7/\mathbb{Z}_k$ in a particular limit.

Let us finally make some considerations regarding the supersymmetries preserved by our brane solution. Even if the 11d metric in equation (\ref{brane_metric_M2M5KKM5_branesol}) is invariant under 
$\mathrm{SO}(3) \times\mathrm{SO}(3)$, the ansatz taken for our branes, which are smeared on the ${\tilde S}^3$, reduces the global symmetries to just the $\mathrm{SO}(3)$ associated to the $S^2$ contained in the worldvolume of the M5'-branes\footnote{Our construction is thus essentially different from the brane set-up that would give rise to the solutions constructed in \cite{DHoker:2008lup,DHoker:2008rje}, in which the branes must be localised on the two 3-spheres.}. This is manifest in the $G_{(4)}$ 4-form flux given by equation (\ref{G4}). The preserved $\mathrm{SO}(3)$ is then the R-symmetry group associated to our solutions, which are, by construction, $\mathcal{N}=(0,4)$ supersymmetric. Regarding the introduction of the two families of KK-monopoles, one can check by studying the supersymmetry projectors of the brane solution that the introduction of one of the two types is for free, in the sense that it does not reduce further the supersymmetries preserved by the rest of the branes. One can see explicitly that this happens thanks to the presence of the M2-branes in the background. 

\subsection{Surface defects as 7d charged domain walls}
\label{7dDWchange}

We can now show that the $\mrm{AdS}_3$ background \eqref{brane_metric_M2M5KKM5_nh} admits, in a particular limit, a local description in terms of the $\mathrm{AdS}_7/\mathbb{Z}_k$ vacuum of M-theory. The idea is to relate the near-horizon geometry \eqref{brane_metric_M2M5KKM5_nh} to a charged 7d domain wall characterised by an $\mrm{AdS}_3$ slicing and an asymptotic behaviour that reproduces locally the $\mathrm{AdS}_7$ vacuum of $\ma N=1$ 7d supergravity. The reason the vacuum appears asymptotically locally is that the presence of the M2-M5 defect breaks its isometries (this is most manifest by the non-vanishing 4-form flux), as well as half of its supersymmetries. 

We start considering $\ma N=1$ minimal gauged supergravity in seven dimensions and its embedding in M-theory, as outlined in appendix \ref{7dsugra}. In this case the minimal field content (excluding the presence of vectors) is given by the gravitational field, a real scalar $X_7$ and a 3-form gauge potential $\ma B_{(3)}$. The 7d background in which we are interested was introduced in~\cite{Dibitetto:2017tve} and further studied in~\cite{Dibitetto:2017klx}. It has the following form
\begin{equation}
\begin{split}\label{7dAdS3}
& ds^2_7=e^{2U(\mu)}\left(ds^2_{\text{AdS}_3}+ds^2_{S^3} \right)+e^{2V(\mu)}d\mu^2\,,\\
&\ma B_{(3)}=b(\mu)\,\left(\text{vol}_{\text{AdS}_3}+\text{vol}_{S^3}\right)\,,\\
&X_7=X_7(\mu)\,.
\end{split}
\end{equation}
The BPS equations were worked out in \cite{Dibitetto:2017tve} and are given by
\begin{equation}
 \begin{split}
  U^\prime= \frac{2}{5}\,e^{V}\,f_7\,,\qquad X_7^\prime=-\frac{2}{5}\,e^{V}\,X_7^2\,D_Xf_7\,,\qquad b^\prime=- \frac{2\,e^{2U+V}}{X_7^2}\,.
  \label{chargedDW7d}
 \end{split}
\end{equation}
In these equations $f_7$ is the superpotential, defined in \eqref{7dsuperpotential}. The flow \eqref{chargedDW7d} preserves 8 real supercharges (it is BPS/2 in 7d). In order to be consistent it has to be endowed by the odd-dimensional self-duality condition  \eqref{odddimselfdual}. This relation takes the form
\begin{equation}  \label{chargedDW7d1}
b=-\frac{e^{2U}\,X_7^2}{h}\,.
\end{equation}
We can work out an explicit solution by choosing a gauge,
\begin{equation}
 e^{-V}=-\frac25\,X_7^2\,D_Xf_7\,,
\end{equation}
such that system \eqref{chargedDW7d} can be easily integrated to give~\cite{Dibitetto:2017tve}
\begin{equation}
 \begin{split}
  e^{2U}= &\ 2^{-1/4}g^{-1/2}\,\left(\frac{\mu}{1-\mu^5}\right)^{1/2}\ , \qquad e^{2V}=\frac{25}{2\,g^2}\, 
  \frac{\mu^6}{\left(1- \mu^5\right)^2}\ ,\\
   b=&\ -2^{1/4}\,g^{-3/2}\,\frac{\mu^{5/2}}{(1-\mu^5)^{1/2}}\ ,\qquad \ X_7=\mu\ ,
   \label{chargedDWsol7d}
 \end{split}
\end{equation}
with $\mu$ running between 0 and 1 and $h= \frac{g}{2\sqrt2}$. The behaviour at the boundaries is such that when  $\mu \rightarrow 1$ the domain wall \eqref{7dAdS3} is locally $\mathrm{AdS}_7$, since we have
\begin{equation}
 \begin{split}
  \ma {R}_{7}= -\frac{21}{4}\,g^2+\ma O (1-\mu)^{2}\,,\qquad X_7=&\ 1+\ma O (1-\mu)\ ,
  \label{UVchargedDW7d}
 \end{split}
\end{equation}
where $\mathcal{R}_{7}$ is the 7d scalar curvature. In turn, when $\mu \rightarrow 0$ the 7d spacetime exhibits a singular behaviour. 
We point out that the background \eqref{7dAdS3} can be generalised by quotienting the 3-sphere (locally written as in \eqref{orbifoldS3}) without any further breaking of the supersymmetries, i.e. $ds^2_{S^3}\rightarrow ds^2_{S^3/\mathbb{Z}_k}$ and $\text{vol}_{S^3}\rightarrow \text{vol}_{S^3/\mathbb{Z}_k}$.

The uplift of the 7d background to M-theory takes place using the relations \eqref{truncationansatz7d} and \eqref{truncationansatz7dfluxes}, summarised in appendix \ref{7dsugra}. This gives
\begin{equation}
 \begin{split}\label{uplift7dDW}
  ds^2_{11}&=\Sigma_7^{1/3}\,e^{2U}\left(ds^2_{{\scriptsize \mrm{AdS}_3}}+ds^2_{S^3/\mathbb{Z}_k} \right)+\Sigma_7^{1/3}e^{2V}d\mu^2\\
  &+2g^{-2}\Sigma_7^{1/3}\,X_7^{3}\,d\xi^2+2g^{-2}\,X_7^{-1}\,\Sigma_7^{-2/3}\,c^{2}\,ds^2_{\tilde S^3}\ ,\\
 G_{(4)}&=\left(s\,b^\prime\,d\mu + c\, b\,d\xi \right)\wedge \text{vol}_{{\scriptsize \mrm{AdS}_3}}+\left(s\,b^\prime\,d\mu + c\, b\,d\xi \right)\wedge \text{vol}_{S^3/\mathbb{Z}_k}\\
 &-\frac{4}{\sqrt 2}\,g^{-3}\,c^{3}\,\Sigma_7^{-2}\,W\,d\xi\,\wedge\,\text{vol}_{\tilde S^3}-\frac{20}{\sqrt{2}}\,g^{-3}\,\Sigma_7^{-2}\,X_7^{-4}\,s\,c^4\,X_7^\prime\,d\mu\wedge\,\text{vol}_{\tilde S^3}\,,\\
 \end{split}
\end{equation}
where $c=\cos\xi$, $s=\sin \xi\,,\,\,\Sigma_7=X_7\,c^2+X_7^{-4}\,s^2$ and $W$ is given by \eqref{truncationansatz7dfluxes}. We can now relate this solution to the near horizon geometry given by equation \eqref{brane_metric_M2M5KKM5_nh}. We consider for simplicity a round $\tilde S^3$. This can be immediately generalised to the case in which KK'-monopoles are included by modding out the $\tilde S^3$.

One can see that the near-horizon geometry~\eqref{brane_metric_M2M5KKM5_nh} takes the form given in~\eqref{uplift7dDW} if one
redefines the $(z, \rho)$ coordinates in terms of the ``domain wall coordinates" $(\mu, \xi)$ as
\begin{equation}\label{coord7dAdS7}
 z=\frac{\sqrt 2}{4g\,k\, Q_{\mathrm{M}5}}\,\sin\xi\,e^{2U}\,X_7^2\,, \qquad \rho=\frac{\sqrt 2}{4\,g\,k\, Q_{\mathrm{M}5}}\,\cos\xi\,e^{2U}\,X_7^{-1/2}\,,
\end{equation}
and requires that
\begin{equation}\label{H5sol}
  H_{\mathrm{M}5'}=\frac{2^6Q_{\mathrm{M}5}^3\,k^3\,e^{-6U}}{\Sigma_7}\,.
\end{equation}
In this calculation one needs to crucially use the 7d BPS equations \eqref{chargedDW7d} and the self-duality condition  \eqref{chargedDW7d1}. The expression for $H_{\mathrm{M}5'}$ given by equation (\ref{H5sol}) satisfies the condition imposed by equation \eqref{11d-motherbranesEOM}. 
The $\mathrm{AdS}_7$ geometry arises through a non-linear change of coordinates that relates the $(z, \rho)$ coordinates of the near horizon $\mathrm{AdS}_3$ geometry to the 
$(\mu, \xi)$ coordinates of the 7d domain wall solution, in which the defect interpretation becomes manifest. When $\mu\to 1$ the domain wall reaches locally the $\mrm{AdS}_7/\mathbb{Z}_k$ vacuum, while, entering into the 7d bulk, the isometries of the vacuum are broken by the $\mrm{AdS}_3$ slicing and 3-form gauge potential, that capture the effects produced by the M2-M5 brane intersection. This allows us to interpret the singular behaviour appearing in 7d when $\mu\rightarrow 0$ in terms of M2-M5 brane sources.

Finally we point out that the choice of the coordinates $(\mu, \xi)$ allows to describe holographically the location of the defect, by just studying the boundary metric of the 7d domain wall \eqref{7dAdS3}. This argument was originally presented in \cite{Clark:2004sb} and it was applied to the domain wall given by \eqref{7dAdS3} in \cite{Dibitetto:2017klx}. Writing $ds^2_{{\scriptsize \mrm{AdS}_3}}=\zeta^{-2}(d\zeta^2+ds^2_{\mathbb{R}^{1,1}})$, it is easy to see that the metric in the $(\zeta, R)$- plane, with $dR=e^{V-U}d\mu$, has a conical defect at $\zeta=0$. This fixes the position of the defect and allows to interpret the $\mu$ coordinate as an angular coordinate defining the wedge in which a 7d observer probes the defect geometry.

\section{Surface defects in massless IIA}
\label{IIApicture}

\begin{figure}[http!]
\begin{center}
\scalebox{1}[1]{ \xymatrix@C-6pc {\text{  } & *+[F-,]{\begin{array}{c} \textrm{M2 - M5  on KK  - M5' - KK'}\vspace{2mm} \\ \textrm{AdS}_3\times S^3/\mathbb{Z}_k  \times \tilde{S}^3/\mathbb{Z}_{k'} \times I_\rho \times I_z \\ \subset \mathrm{AdS}_7/\mathbb{Z}_k \times S^4/\mathbb{Z}_{k'} \end{array}} \ar[dl]^\phi\ar[dr]^\chi & \text{  }   \\ 
*+[F-,]{\begin{array}{c} \textrm{D2 - D4 on KK - NS5 - D6} \vspace{2mm} \\ \textrm{AdS}_3\times S^3/\mathbb{Z}_k  \times \tilde{S}^2\times I_\rho \times I_z \\ \subset \mathrm{AdS}_7/\mathbb{Z}_k \times \tilde S^2 \times I \end{array}} &  \text{  }  & *+[F-,]{\begin{array}{c} \textrm{ D2 - NS5 - D6 on D4 - KK' } \vspace{2mm} \\ \textrm{AdS}_3\times S^2  \times \tilde S^3/\mathbb{Z}_{k'}\times I_\rho \times I_z\end{array}}
 }}
\end{center}
\caption{Reductions of the KK'-M5'-M2-M5-KK brane system to Type IIA and their near-horizon limits. Only the reduction along $\phi$ asymptotes to $\mrm{AdS}_7$, with the KK-M5'-KK' system becoming KK-NS5-D6.}\label{fig} 
\end{figure}
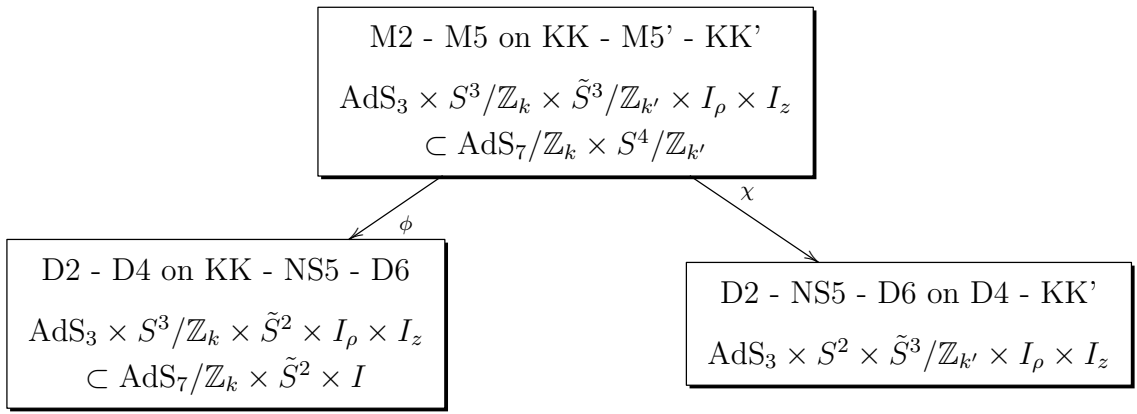

In this section we study the Type IIA regime of the M-theory set-up introduced in the previous section. From a 10d point of view the KK'-M5'-M2-M5-KK system has two different descriptions, depending on whether the reduction is performed on a circle that lies inside or outside the worldvolume of the M5'-branes.
We recall that the 11d background has two compact coordinates. The $\chi$ coordinate lies inside the worldvolume of the M5'-branes and is identified as the Taub-NUT direction of the KK-monopoles. In turn, the $\phi$ coordinate lies outside the worldvolume of the M5'-branes and is identified as the Taub-NUT direction of the KK'-monopoles. The two possible reductions to Type IIA are depicted in Figure \ref{fig}.

In 10d one observes an interesting phenomenon. Both reductions produce a D2-D4-NS5-D6 intersection with Kaluza-Klein monopoles, and both of them are described by near-horizon geometries with the same topology and supersymmetries. The charge distributions of the branes are however essentially different. In the first reduction the $\mathrm{AdS}_3$ near-horizon geometries constitute a new class of solutions to massless Type IIA, that we will further explore in this paper. These solutions   enjoy an interesting defect interpretation in terms of KK-NS5-D6 bound states, dual to an $\mathrm{AdS}_7$ geometry, on which D2-D4 branes end. In the second reduction the ${\tilde S}^3/\mathbb{Z}_{k'}$ and $I_\rho$ sub-manifolds give rise to $\mathbb{C}^2/\mathbb{Z}_{k'}$, such that the resulting $\mathrm{AdS}_3$ near-horizon geometries become the class I family of solutions to Type IIA recently classified in  \cite{Lozano:2019emq}, restricted to the massless case, $\mathrm{CY}_2=\mathbb{C}^2/\mathbb{Z}_{k'}$, $u^\prime=0$ and $H_2=0$ (see  \cite{Lozano:2019emq}). We will see in section \ref{massiveIIA} that these solutions need to be embedded in massive IIA in order to be given a defect interpretation in terms of  D4-KK'-D8 branes on which D2-NS5-D6 branes end.
 Roughly speaking, one could say that in both classes of solutions the D4 and NS5 branes exchange their ``roles", together with the D6-branes and the Kaluza-Klein monopoles. Work in progress shows that the two families of solutions are in fact related upon a chain of T-S-T- dualities \cite{FLP}.


In the remainder of this section we focus on the first reduction, which is the one that preserves the $\mrm{AdS}_7$ asymptotics in the UV. We present the brane picture and show that the resulting near-horizon geometries constitute a new class of $\mrm{AdS}_3$ solutions to Type IIA supergravity with ${\mathcal N}=(0,4)$ supersymmetries. The special feature of this class of solutions, as compared to the solutions in  \cite{Lozano:2019emq}, is that they asymptote (locally) to the $\mathrm{AdS}_7/\mathbb{Z}_k\times S^2\times I$ solution to massless IIA supergravity, and can thus be interpreted as surface defect CFTs within the 6d (1,0) CFT dual to this solution. In section \ref{massiveIIA} we focus on the second reduction. We show that once generalised to massive IIA the solutions describe surface defect CFTs within the 5d fixed point theory dual to the $\mathrm{AdS}_6$ solution of Brandhuber-Oz \cite{Brandhuber:1999np} (with extra KK'-monopoles).

\subsection{New $\mrm{AdS}_3$ solutions with ${\mathcal N}=(0,4)$ supersymmetries}
\label{masslessIIAnh}
In this section we consider the reduction of the 11d background \eqref{brane_metric_M2M5KKM5_branesol} along the Taub-NUT coordinate $\phi$. The resulting Type IIA configuration, depicted in Table  \ref{Table:branesinmasslessIIA}, consists on  D2-D4 branes, coming from the smeared M2-M5 brane system appearing in \eqref{brane_metric_M2M5KKM5_branesol}, ending on a KK-NS5-D6 bound state, that arises upon reduction of the KK-M5'-KK' brane system. As already shown in the literature (see for example \cite{Cvetic:2000cj}) this bound state is described in  the near-horizon limit by an $\mrm{AdS}_7/\mathbb{Z}_k$ vacuum preserving 16 supercharges and a 3d internal space given by a 2-sphere foliation over a segment.
\begin{table}[h!]
\renewcommand{\arraystretch}{1}
\begin{center}
\scalebox{1}[1]{
\begin{tabular}{c||c c|c c c c | c | c c c}
branes & $t$ & $x^1$ & $r$ & $\theta^{1}$ & $\theta^{2}$ & $\chi$ & $z$ & $\rho$ & $\varphi^1$ & $\varphi^2$ \\
\hline \hline
$\mrm{D}6$ & $\times$ & $\times$ & $\times$ & $\times$ & $\times$ & $\times$ & $\times$ & $-$ & $-$ & $-$  \\
$\mrm{NS}5$ & $\times$ & $\times$ & $\times$ & $\times$& $\times$  & $\times$ & $-$ & $-$ & $-$ & $-$  \\
$\mrm{KK}$ & $\times$ & $\times$ & $-$ & $-$ & $-$ & $\mrm{ISO}$ & $\times$ & $\times$ & $\times$ & $\times$  \\
$\mrm{D}2$ & $\times$ & $\times$ & $-$ & $-$ & $-$ & $-$ &  $\times$ & $-$ & $-$ & $-$  \\
$\mrm{D}4$ & $\times$ & $\times$ &$-$ & $-$ & $-$ & $-$ & $-$ & $\times$ & $\times$ & $\times$  \\
\end{tabular}
}
\end{center}
\caption{Brane picture underlying the D2-D4 branes ending on the NS5-D6-KK intersection. The system is $\mrm{BPS}/8$.} \label{Table:branesinmasslessIIA}
\end{table}
We now add the D2-D4 branes to this system. We introduce firstly the 10d metric
\begin{equation}
\label{brane_metric_D2D4KKNS5D6}
\begin{split}
d s_{10}^2&=\,H_{\mathrm{D}6}^{-1/2}\,\left[H_{\mathrm{D}4}^{-1/2}\,H_{\mathrm{D}2}^{-1/2}\,ds^2_{\mathbb{R}^{1,1}}+H_{\mathrm{D}4}^{1/2}\,H_{\mathrm{D}2}^{1/2} \,\left(H_{\text{KK}}(dr^2+r^2d s^2_{S^2})+H_{\text{KK}}^{-1}(d\chi+Q_{\mathrm{KK}}\,\omega)^2\right)\right] \\
&+H_{\mathrm{D}6}^{-1/2}\,H_{\mathrm{NS}5}\,H_{\mathrm{D}4}^{1/2}\,H_{\mathrm{D}2}^{-1/2}\,dz^2+H_{\mathrm{D}6}^{1/2}\,H_{\mathrm{NS}5}\,H_{\mathrm{D}4}^{-1/2}\,H_{\mathrm{D}2}^{1/2}(d\rho^2+\rho^2 ds^2_{\tilde S^2}) \, ,
\end{split}
\end{equation}
where we take the D4 and D2 charges completely localised within the worldvolume of the NS5 branes, i.e. $H_{\mathrm{D}4}=H_{\mathrm{D}4}(r)$ and $H_{\mathrm{D}2}=H_{\mathrm{D}2}(r)$. Secondly, we introduce the following gauge potentials and dilaton,
\begin{equation}
\begin{split}\label{brane_potentials_D2D4NS5D6KK}
&C_{(3)}=H_{\mathrm{D}2}^{-1}\,\text{vol}_{\mathbb{R}^{1,1}}\wedge dz\,,\\
&C_{(5)}=H_{\mathrm{D}6}\,H_{\mathrm{NS}5}\,H_{\mathrm{D}4}^{-1}\,\rho^2\,\text{vol}_{\mathbb{R}^{1,1}}\wedge d\rho \wedge \text{vol}_{\tilde S^2}\,,\\
&C_{(7)}=H_{\mathrm{KK}}\,H_{\mathrm{D}4}\,H_{\mathrm{D}6}^{-1}\,r^2\,\text{vol}_{\mathbb{R}^{1,1}}\wedge dr \wedge \text{vol}_{S^2}\wedge d\chi \wedge dz \,,\\
&B_{(6)}=H_{\mathrm{KK}}\,H_{\mathrm{D}4}\,H_{\mathrm{NS}5}^{-1}\,r^2\,\text{vol}_{\mathbb{R}^{1,1}}\wedge  d r \wedge\text{vol}_{S^2}\wedge d\chi\,,\\ \vspace{0.4cm}
&e^{\Phi}=H_{\mathrm{D}6}^{-3/4}\,H_{\mathrm{NS}5}^{1/2}\,H_{\mathrm{D}2}^{1/4}\,H_{\mathrm{D}4}^{-1/4}\,,
\end{split}
\end{equation}
where we take the NS5-D6 branes completely localised in their transverse space. From \eqref{brane_potentials_D2D4NS5D6KK} one can deduce\footnote{We use the conventions for fluxes of \cite{Imamura:2001cr}. 
  }
the fluxes 
\begin{equation}
\begin{split}\label{fluxes_D2D4NS5D6KK}
& F_{(2)} = -\partial_\rho H_{\mathrm{D}6} \,\rho^2\,\text{vol}_{\tilde S^2}\,, \\
& H_{(3)} = -\partial_\rho H_{\mathrm{NS}5} \, \rho^2\,dz\wedge\text{vol}_{\tilde S^2}+H_{\mathrm{D}2}\,H_{\mathrm{D}4}^{-1}\,H_{\mathrm{D}6}\,\partial_z H_{\mathrm{NS}5}\,\rho^2\,d\rho\wedge\text{vol}_{\tilde S^2}\,, \\
 &F_{(4)}=\partial_rH_{\mathrm{D}2}^{-1}\,\text{vol}_{\mathbb{R}^{1,1}}\wedge dr\wedge dz-\partial_r H_{\mathrm{D}4}\,r^2\, \text{vol}_{S^2}\wedge d\chi\wedge dz\,.
 \end{split}
\end{equation}
As in the 11d picture, the equations of motion and Bianchi identities for the D2-D4-KK branes and the NS5-D6 branes can be solved independently. We have that
\begin{equation}\label{10d-defectbranesEOM}
H_{\mathrm{D}2}=H_{\mathrm{D}4}\,,\qquad \nabla^2_{\mathbb{R}^3_r}\,H_{\mathrm{D}4}=0\qquad \text{with}\qquad H_{\mathrm{KK}}=\frac{Q_{\mathrm{KK}}}{r}\,,
\end{equation}
and for the NS5-D6 branes,
\begin{equation}\label{10d-motherbranesEOM}
\nabla^2_{\mathbb{R}^3_\rho} H_{\mathrm{NS}5} + H_{\mathrm{D}6} \, \partial_z^2 H_{\mathrm{NS}5}=0  \qquad  \text{and}  \qquad  \nabla^2_{\mathbb{R}^3_\rho} H_{\mathrm{D}6} = 0  \,.
\end{equation}
We note that the equations in \eqref{10d-motherbranesEOM} coincide with those found in  \cite{Imamura:2001cr} for the NS5-D6 bound state in the massless limit.
The equations in \eqref{10d-defectbranesEOM} can be easily solved for
\begin{equation}
  H_{\mathrm{D}4}(r)=H_{\mathrm{D}2}(r)=1+\frac{Q_{\mathrm{D}4}}{r}\,,
 \end{equation}
where we introduced the D2 and D4 charges $Q_{\mathrm{D}2}$ and $Q_{\mathrm{D}4}$ that in order to satisfy \eqref{10d-defectbranesEOM} have to be equal. We point out that uplifting to 11d we get the background \eqref{brane_metric_M2M5KKM5_branesol} with $Q_{\mathrm{D}2}=Q_{\mathrm{M}2}$, $Q_{\mathrm{D}4}=Q_{\mathrm{M}5}$, $H_{\text{D}6} = H_{\text{KK}'}/4$ and a rescaling $\rho\to 2\rho$ in the 10d solution.

We now analyse the limit $r \rightarrow 0$. As we already saw in the 11d case, the KK-monopole charge $ Q_{\mathrm{KK}}=k$ placed on the worldvolume of the NS5-branes realises the orbifolded 3-sphere $S^3/\mathbb{Z}_k$. The metric \eqref{brane_metric_D2D4KKNS5D6} and the fluxes \eqref{fluxes_D2D4NS5D6KK} take the form\footnote{We redefined the Minkowski coordinates as $(t,x^1)\rightarrow 2\,Q_{\mathrm{D}4}\,Q_{\text{KK}}^{1/2}\,(t,x^1)$ and rescaled the function $H_{\text{D}6}\to H_{\text{D}6}/2$.}
\begin{equation}
\label{brane_metric_D2D4KKNS5D6_nh}
\begin{split}
ds_{10}^2 &= 4\sqrt{2} \, k \, Q_{\text{D}4} H_{\mathrm{D}6}^{-1/2} \left[ds^2_{\text{AdS}_3} + ds^2_{S^3/\mathbb{Z}_k} \right] + \sqrt{2} \, H_{\mathrm{D}6}^{-1/2} H_{\mathrm{NS}5} \, dz^2 + \frac{1}{\sqrt{2}} H_{\mathrm{D}6}^{1/2} H_{\mathrm{NS}5} \left(d\rho^2 + \rho^2 ds^2_{\tilde{S}^2}\right) \,, \\
F_{(2)} &= \frac{Q_{\mathrm{D}6}}{2} \, \text{vol}_{\tilde S^2} \,,  \qquad \qquad  e^{\Phi} = 2^{3/4} H_{\mathrm{D}6}^{-3/4} H_{\mathrm{NS}5}^{1/2} \,,\\
H_{(3)} &= -\partial_\rho H_{\mathrm{NS}5} \, \rho^2 \, dz \wedge \text{vol}_{\tilde S^2} + \frac12 \, H_{\mathrm{D}6} \, \partial_z H_{\mathrm{NS}5} \, \rho^2 \, d\rho \wedge \text{vol}_{\tilde S^2} \,, \\
 F_{(4)} &= 8 \, k\,Q_{\text{D}4} \, \text{vol}_{\text{AdS}_3}\wedge dz + 8 \, k \, Q_{\text{D}4} \, \text{vol}_{S^3/\mathbb{Z}_k} \wedge dz \,,\\
\end{split}
\end{equation}
with
\begin{equation}
\label{10d-motherbranesEOM_nh}
\nabla^2_{\mathbb{R}^3_\rho} H_{\mathrm{NS}5} + \frac12 H_{\mathrm{D}6} \,\partial_z^2 H_{\mathrm{NS}5} =0 \qquad  \text{and}  \qquad  H_{\mathrm{D}6} = \frac{Q_{\mathrm{D}6}}{\rho} \,,
\end{equation}
where the D6-brane charge $ Q_{\mathrm{D}6}$ equals the KK' monopole charge of the 11d background \eqref{brane_metric_M2M5KKM5_nh},  $ Q_{\mathrm{D}6}=k'$. 

The $\mathrm{AdS}_3$ backgrounds given by equation (\ref{brane_metric_D2D4KKNS5D6_nh}), with $H_{\mathrm{NS}5}$ and $H_{\mathrm{D}6}$ satisfying \eqref{10d-motherbranesEOM_nh}, constitute a new class of 10d backgrounds with ${\mathcal N}=(0,4)$ supersymmetries. These solutions are of the form AdS$_3\times S^3/\mathbb{Z}_k\times S^2$ fibered over two intervals. They preserve the same number of supersymmetries as the AdS$_3\times S^2\times \mathrm{CY}_2\times I$ solutions constructed in \cite{Lozano:2019emq} and involve the same types of branes (in the massless limit of the solutions in  \cite{Lozano:2019emq}), plus extra KK-monopoles\footnote{That can also be introduced in the AdS$_3\times S^2\times \mathrm{CY}_2\times I$ solutions in  \cite{Lozano:2019emq} without any further breaking of the supersymmetries.}. As mentioned, the brane intersections are however different. 

In appendix \ref{newAdS3} we show that a broader class of $\mathrm{AdS}_3\times S^3/\mathbb{Z}_k\times S^2$ solutions fibered over two intervals and preserving ${\mathcal N}=(0,4)$ supersymmetries can in fact be constructed from the general class of 
$\mathrm{AdS}_3\times S^3/\mathbb{Z}_k \times \mathrm{CY}_2\times I$ solutions to M-theory recently constructed in~\cite{Lozano:2020bxo}. In order to obtain this broader class one needs to take the $\mathrm{CY}_2$ to be $T^4$, or rather $\mathbb{R}^4$, and  reduce  on the Hopf-fibre of the 3-sphere contained in this space. In the remainder of the paper we will however focus our attention on the more restrictive case defined by (\ref{brane_metric_D2D4KKNS5D6_nh}).
In the next section we will relate this solution to a domain wall solution that asymptotes locally to AdS$_7/\mathbb{Z}_k$ and give it an interpretation as dual to D2-D4 surface defects within the corresponding 6d (1,0) dual CFT. 

\subsection{Surface defects within the NS5-D6-KK brane system}

In this section we follow the same strategy of section \ref{7dDWchange} in order to relate the new 
$\mathrm{AdS}_3\times S^3/\mathbb{Z}_k\times S^2$ solutions given by equation (\ref{brane_metric_D2D4KKNS5D6_nh}) to an $\mathrm{AdS}_7$ geometry in the UV. In this case we relate the solutions  to the uplift of the 7d domain wall discussed in section \ref{7dDWchange} to massless IIA supergravity. The 10d domain wall solution flows in the UV to the $\mathrm{AdS}_7\times S^2\times I$ solution to massless IIA supergravity found in \cite{Cvetic:2000cj}, modded by $\mathbb{Z}_k$, which arises in the near-horizon limit of a NS5-D6-KK brane intersection. This solution belongs to
the general class of solutions to massive IIA supergravity constructed in 
\cite{Apruzzi:2013yva}, modded by $\mathbb{Z}_k$, in the massless limit. The solutions to massive IIA supergravity in \cite{Apruzzi:2013yva} are the near horizon geometries of NS5-D6-D8 brane intersections \cite{Bobev:2016phc}, and encode very naturally the information of the 6d (1,0) dual CFTs that live in their worldvolumes \cite{Cremonesi:2015bld}.
For this reason, we will follow the notation in \cite{Apruzzi:2013yva,Cremonesi:2015bld} in this section. In the same vein, we will use the uplift formulae from 7d ${\mathcal N}=1$ supergravity  to massive IIA supergravity found in \cite{Passias:2015gya}, which we will particularise to the massless case. This parametrisation will be very convenient when we discuss the 2d CFTs dual to our solutions in section  \ref{defect-quivers}.

We start recalling the $\mathrm{AdS}_7\times S^2\times I$ solution to massless IIA supergravity of \cite{Cvetic:2000cj} using the parametrisation of \cite{Apruzzi:2013yva}. We then study the 10d domain wall solution that asymptotes locally to this solution and relate it to our solution (\ref{brane_metric_D2D4KKNS5D6_nh}). Finally, we present in section \ref{defect-quivers} the explicit 2d CFT dual to our solution and show that it occurs as a surface defect within the 6d CFT dual to the   $\mathrm{AdS}_7$ solution to massless IIA.

\subsubsection{The AdS$_7/\mathbb{Z}_k$ solution to massless IIA}

The general class of solutions to massive Type IIA supergravity constructed in  \cite{Apruzzi:2013yva}
consists on foliations of AdS$_7\times S^2$ over an interval preserving 16 supersymmetries. 
Using the parametrisation in \cite{Cremonesi:2015bld} they can be completely determined by a function $\alpha(y)$ that satisfies the differential equation\footnote{Note that we use $y$ instead of $z$  as in \cite{Cremonesi:2015bld} in order to avoid confusion with the notation in the previous sections.}
  \begin{equation}
\label{dddotalfa}
\dddot{\alpha}=-162\pi^3 F_{(0)},
\end{equation}
where $F_{(0)}$ is the RR 0-form.  Here we will be concerned with the massless case, for which $\dddot{\alpha}=0$. For $F_{(0)}=0$ the metric and fluxes are given by
\begin{align}
\label{metricAdS7}
ds_{10}^2 &= \pi\sqrt{2} \bigg[ 8 \Bigl(-\frac{\alpha}{\ddot{\alpha}}\Bigr)^{1/2} ds^2_{\text{AdS}_7} + \Bigl(-\frac{\ddot{\alpha}}{\alpha}\Bigr)^{1/2} dy^2 + \Bigl(-\frac{\alpha}{\ddot{\alpha}}\Bigr)^{1/2} \frac{(-\alpha\ddot{\alpha})}{\dot{\alpha}^2 - 2\alpha\ddot{\alpha}} ds^2_{S^2} \bigg] \,, \\
\label{dilatonAdS7}
e^{2\Phi} &= 3^8 2^{5/2} \pi^5 \frac{(-\alpha/\ddot{\alpha})^{3/2}}{\dot{\alpha}^2 - 2\alpha\ddot{\alpha}}\,, \\
\label{B2AdS7}
B_{(2)} &= \pi \Bigl(-y + \frac{\alpha\dot{\alpha}}{\dot{\alpha}^2 - 2\alpha\ddot{\alpha}}\Bigr) \, \text{vol}_{S^2} \,, \\
\label{F2AdS7}
F_{(2)} &= -\frac{\ddot{\alpha}}{162\pi^2} \, \text{vol}_{S^2} \,.
\end{align}
In the most general case in which  $F_{(0)}\neq 0$ the backgrounds in \cite{Apruzzi:2013yva}
 arise as near horizon geometries of D6-NS5-D8 brane intersections, from which 6d linear quivers with 8 supercharges can be explicitly constructed~\cite{Gaiotto:2014lca,Cremonesi:2015bld}. In these brane set-ups the NS5-branes are located at fixed positions in $y$, the D6-branes are stretched between them in this direction and the D8-branes are perpendicular.  In the massless case
 we will take
\begin{equation}
\label{alphaz}
\alpha(y)=-\frac12 \alpha_0 y^2 + \beta_0 y\, \qquad \Rightarrow \qquad \ddot{\alpha}=-\alpha_0\, ,
\end{equation}
with $\alpha_0, \beta_0>0$,
such that the space is terminated by D6-branes at both ends of the $y$-interval, $y=0$ and $y=2\beta_0/\alpha_0$. 
The solution arises as the near-horizon geometry of the D6-NS5 brane intersection depicted in Table \ref{6dmassless} \cite{Cvetic:2000cj,Bobev:2016phc}. In M-theory it involves M5-branes intersected with KK-monopoles, which render the 6d CFT living in the M5-branes (1,0) supersymmetric. One can check that it is possible to add a second stack of $k$ KK-monopoles, modding out the AdS$_7$ subspace to AdS$_7/\mathbb{Z}_k$, without breaking any further supersymmetry. The resulting brane intersection in Type IIA is depicted in Table \ref{6dmassless2}.
\begin{table}[h!]
\renewcommand{\arraystretch}{1}
\begin{center}
\scalebox{1}[1]{
\begin{tabular}{c||c c c c c c | c | c c c}
branes & $t$ & $x^1$ & $x^2$ & $x^3$ & $x^4$ & $x^5$ & $y$ & $\rho$ & $\varphi^1$ & $\varphi^2$ \\
\hline \hline
$\mrm{D}6$ & $\times$ & $\times$ & $\times$ & $\times$ & $\times$ & $\times$ & $\times$ & $-$ & $-$ & $-$  \\
$\mrm{NS}5$ & $\times$ & $\times$ & $\times$ & $\times$& $\times$  & $\times$ & $-$ & $-$ & $-$ & $-$  \\
\end{tabular}
}
\end{center}
\caption{$\frac14$-BPS brane intersection underlying the massless AdS$_7$ solution to Type IIA.  The 6d (1,0) dual CFT lives in the $(t,x^1,x^2,x^3,x^4,x^5)$ directions. $y$ is the field theory direction.}  
	\label{6dmassless}
\end{table}

\begin{table}[h!]
\renewcommand{\arraystretch}{1}
\begin{center}
\scalebox{1}[1]{
\begin{tabular}{c||c c c c c c | c | c c c}
branes & $t$ & $x^1$ & $x^2$ & $x^3$ & $x^4$ & $x^5$ & $y$ & $\rho$ & $\varphi^1$ & $\varphi^2$ \\
\hline \hline
$\mrm{D}6$ & $\times$ & $\times$ & $\times$ & $\times$ & $\times$ & $\times$ & $\times$ & $-$ & $-$ & $-$  \\
$\mrm{NS}5$ & $\times$ & $\times$ & $\times$ & $\times$& $\times$  & $\times$ & $-$ & $-$ & $-$ & $-$  \\
$\mrm{KK}$ & $\times $ & $\times $ & $-$ & $-$ & $-$ & $\mrm{ISO}$ & $\times$ & $\times$ & $\times$ & $\times$ \\
\end{tabular}
}
\end{center}
\caption{$\frac14$-BPS brane intersection underlying the massless AdS$_7/\mathbb{Z}_k$ solution to Type IIA.  The 6d (1,0) dual CFT lives in the $(t,x^1,x^2,x^3,x^4,x^5)$ directions. $x^5$ is the Taub-NUT direction of the KK-monopoles.  $y$ is the field theory direction.}  
	\label{6dmassless2}
\end{table}

\begin{figure}[h!]
\centering
\includegraphics[scale=0.8]{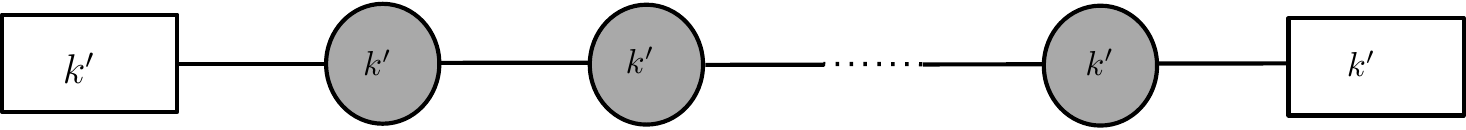}
\caption{6d quiver CFT dual to the AdS$_7/\mathbb{Z}_k$ solution to massless Type IIA.}
\label{6dquiver}
\end{figure}

The 6d quiver CFT dual to the solution can be easily read from the $Q_{\mathrm{D}6}$ and $Q_{\mathrm{NS}5}$ quantised charges,
\begin{eqnarray}
Q_{\mathrm{D}6}&=&\frac{1}{2\pi}\int_{S^2}F_{(2)}=\frac{\alpha_0}{81\pi^2}\,, \label{Q6AdS7}\\
Q_{\mathrm{NS}5}&=&\frac{1}{4\pi^2}\int_{I_y\times S^2}H_{(3)}=\frac{2\beta_0}{\alpha_0}. \label{NS5AdS7}
\end{eqnarray}
These expressions fix $\alpha_0$, $\beta_0$ in terms of the numbers of D6 and NS5 branes of the solution. They show that there are $Q_{\mathrm{NS}5}-1$ stacks of $Q_{\mathrm{D}6}$ D6-branes stretched between $Q_{\mathrm{NS}5}$ parallel NS5-branes, located at $y=1,2,\dots, 2\beta_0/\alpha_0$. Extra D6-branes at both ends provide for the additional $Q_{\mathrm{D}6}$ flavour groups that are required by anomaly cancellation. The resulting 6d (1,0) quiver CFT dual to the solution is depicted in Figure \ref{6dquiver}, where we have used that $Q_{\mathrm{D}6}=k'$.

\subsubsection{AdS$_3\times S^3/\mathbb{Z}_k\times S^2\times I$ asymptotically locally AdS$_7/\mathbb{Z}_k\times S^2$}

In this section we uplift the 7d domain wall solution presented in section \ref{7dDWchange} to 10d, using the uplift formulas to massive IIA supergravity constructed in \cite{Passias:2015gya}, that we truncate  to the massless case. This will be the most adequate framework for the holographic study that we will perform in the next section. The uplift formulas read 
\begin{align}
\label{metricAdS7X}
ds_{10}^2 &= \frac{16\pi}{g} \Bigl(-\frac{\alpha}{\ddot{\alpha}}\Bigr)^{1/2} X_7^{-1/2} ds^2_7 + \frac{16\pi}{g^3} X_7^{5/2} \biggl[ \Bigl(-\frac{\ddot{\alpha}}{\alpha}\Bigr)^{1/2} dy^2 +\Bigl(-\frac{\alpha}{\ddot{\alpha}}\Bigr)^{1/2} \frac{(-\alpha \ddot{\alpha})}{\dot{\alpha}^2-2\alpha\ddot{\alpha} X_7^5} ds^2_{S^2} \bigg] \,, \\
\label{dilatonAdS7X}
e^{2\Phi} &= \frac{3^8 2^6 \pi^5}{g^3} \frac{X_7^{5/2}}{\dot{\alpha}^2-2\alpha\ddot{\alpha} X_7^5} \Bigl(-\frac{\alpha}{\ddot{\alpha}}\Bigr)^{3/2} \,,\\
\label{B2AdS7X}
B_{(2)} &= \frac{2^3 \sqrt{2} \pi}{g^3} \biggl( -y + \frac{\alpha\dot{\alpha}}{\dot{\alpha}^2-2\alpha\ddot{\alpha} X_7^5} \biggr) \, \text{vol}_{S^2}\,, \\
\label{F2AdS7X}
F_{(2)} &= - \frac{\ddot{\alpha}}{162 \pi^2} \, \text{vol}_{S^2}\,, \\
\label{F4AdS7X}
F_{(4)} &= \frac{2^3}{3^4 \pi} \bigl(\ddot{\alpha} \, dy \wedge \mathcal{B}_{(3)} + \dot{\alpha} \, d\mathcal{B}_{(3)}\bigr) \,, \\
F_{(6)} & = \frac{2^8}{3^4 g^4}\frac{(-\alpha\ddot{\alpha})X_7^{2}\,e^{2U}}{\dot{\alpha}^2-2\alpha\ddot{\alpha}X_7^5}\bigl(\sqrt{2}\, g\, e^V \, \alpha\, X_7\, d\mu+\dot{\alpha}\,dy\bigr)\wedge 
\bigl(\text{vol}_{\text{AdS}_3}+\text{vol}_{S^3/\mathbb{Z}_k}\bigr)\wedge \text{vol}_{S^2}\,,
\label{F6AdS7X}
\end{align}
where $ds_7^2$, $X_7$ and $\mathcal{B}_{(3)}$ are the 7d fields defined in (\ref{7dAdS3}).
This solution asymptotes locally when $\mu\rightarrow 1$ to the AdS$_7$ solution summarised in the previous section, given by equations (\ref{metricAdS7})-(\ref{F2AdS7}), for $g^3=2^{7/2}$. In turn, when $\mu \rightarrow 0$ it exhibits a singular behaviour. 

We can now relate the previous domain wall solution to the AdS$_3\times S^3/\mathbb{Z}_k\times S^2$ solution defined by equation (\ref{brane_metric_D2D4KKNS5D6_nh}). 
The near horizon geometry~\eqref{brane_metric_D2D4KKNS5D6_nh} takes the form given by~\eqref{metricAdS7X}-\eqref{F4AdS7X} if one redefines the $(z,\rho)$ coordinates in terms of the domain wall coordinates $(\mu,y)$ as
\begin{equation}
\label{changeofcoord}
z= -\frac{1}{3^4 \pi k \, Q_{\text{D}4}} \, \dot{\alpha} \, b \,,  \qquad  \rho = \frac{8}{3^4 g^2 k^2 Q_{\text{D}4}^2} \, \alpha \, X_7^{-1} e^{4U}\, ,
\end{equation}
and requires that 
\begin{equation}
\label{hNS5}
Q_{\text{D}6} = \frac{(-\ddot{\alpha})}{81 \pi^2} \,,  \qquad  H_{\mathrm{NS}5} = 3^8 \pi^2 k^3 Q_{\text{D}4}^3 \frac{X_7^4 e^{-6U}}{\dot{\alpha}^2 - 2\alpha \ddot{\alpha} X_7^5} \,.
\end{equation}
In this calculation one needs to crucially use the 7d BPS equations~\eqref{chargedDW7d} and the self-duality condition~\eqref{chargedDW7d1}, and take $h=\frac{g}{2\sqrt{2}}$. Further, the $S^3$ in the 7d background (\ref{7dAdS3}) must be modded by $\mathbb{Z}_k$.
The first condition in (\ref{hNS5}) shows that $\alpha_0$ is again fixed by the number of D6-branes of the solution, as in (\ref{Q6AdS7}). 
The second condition is the 10d version of (\ref{H5sol}). In this case one can see that the constraint on $H_{\mathrm{NS}5}$ in~\eqref{10d-motherbranesEOM} is satisfied by means of the BPS equations for $X_7$ and $U$. 
Note that given (\ref{alphaz}) it is enough to take $y\in [0,\beta_0/\alpha_0]$ in order to cover the $z\in [0, \infty)$, $\rho \in [0, \infty)$  intervals of the
AdS$_3\times S^3/\mathbb{Z}_k\times S^2$ solution. However, we are interested in embedding the AdS$_3$ solution into AdS$_7$ also globally. For that purpose the $I_y$ space of the AdS$_3$ solution must also be terminated by D6-branes at both ends of the interval. In order to achieve this we consider two copies of the solution, glued at $z=0$, through
\begin{equation}
z= -\frac{1}{3^4 \pi k \, Q_{\text{D}4}} \, |\dot{\alpha}| \, b\, .
\end{equation}
This allows us to identify $y$ as the field theory direction, by analogy with the role it plays in the AdS$_7$ solution. $\mu$ is identified in turn as the energy scale, as in the domain wall solution.
 
 The second condition in (\ref{hNS5})  singles out a particular solution in the class defined by  (\ref{brane_metric_D2D4KKNS5D6_nh}) that asymptotes locally to the AdS$_7/\mathbb{Z}_k\times S^2$ vacuum of massless 10d supergravity. The AdS$_7$ geometry arises through a non-linear change of variables, that relates the $(z,\rho)$ coordinates of the near horizon AdS$_3$ solution to the $(\mu,y)$ coordinates of the uplifted domain wall solution. In the new coordinates the defect interpretation becomes manifest. When $\mu\rightarrow 1$ the domain wall reaches the AdS$_7/\mathbb{Z}_k$ vacuum, while when $\mu\rightarrow 0$ a singular behaviour describes D2-D4 brane sources, that create a defect when they intersect the NS5-D6-KK brane system, breaking the isometries of $\mrm{AdS}_7$ to those of $\mrm{AdS}_3$. 
 In the next subsection we turn to the construction of its 2d dual CFT.

\subsection{Surface defects CFTs} \label{defect-quivers}

As we have seen, the brane picture associated to the AdS$_7$ solution consists on D6-branes stretched in the $y$ direction between NS5-branes located at fixed positions in $y$, $y=1,2,\dots,\frac{2\beta_0}{\alpha_0}$. At both ends of the interval D6-branes terminate the geometry, and provide the required flavour groups for anomaly cancellation. The associated quiver is depicted in Figure \ref{6dquiver}. In the presence of the defect D2-D4 branes the total  number of NS5-branes does not change, since one can check that the $B_{(2)}$-field does not depend on $X_7$ at either end of the interval. We set $n+1\equiv Q_{\text{NS}5}=2\beta_0/\alpha_0$, and take the NS5-branes positioned at $y=j=1,2,\dots,n+1$. One can check that at each interval 
$y\in [j,j+1]$ a large gauge transformation of gauge parameter $j$ must be performed, such that the condition
$\frac{1}{4\pi^2}\oint_{S^2}B_{(2)}\in [0,1)$ is satisfied. The number of D2-branes stretched between NS5-branes depends on this number, as large gauge transformations contribute to the magnetic component of the RR 6-form Page flux, under which the D2-branes are charged. These, together with the magnetic components of the 4-form RR Page flux, read
\begin{eqnarray}
{\hat F}_{(6)}&=&\frac{2^7}{3^4 g^4}X_7^{-2}\, e^{2U}\Bigl( \sqrt{2} \,g \,e^V \bigl(\alpha - (y-j) \,\dot{\alpha}\bigr) d\mu-2\,y\,\ddot{\alpha} \,X_7^4 \,dy\Bigr)\wedge  \text{vol}_{S^3/\mathbb{Z}_k}\wedge  \text{vol}_{S^2},  \label{F6quiver}\\
{\hat F}_{(4)}&=&\frac{2^{10/3}}{3^4 \pi}\,d(\dot{\alpha}\,X_7^2\, e^{2U})\wedge \text{vol}_{S^3/\mathbb{Z}_k}\, , \label{F4quiver}
\end{eqnarray}
where we have used (\ref{B2AdS7X})-(\ref{F6AdS7X}) together with equations (\ref{7dAdS3}) and (\ref{chargedDW7d1}).

\begin{figure}
\centering
\includegraphics[scale=0.8]{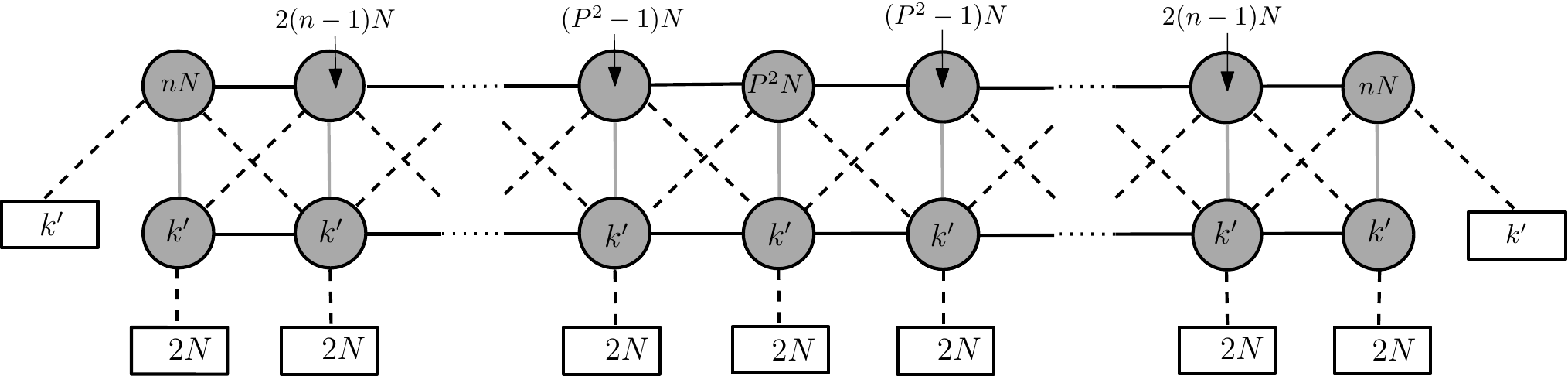}
\caption{2d quiver CFT dual to the AdS$_3\times S^3/\mathbb{Z}_k\times S^2$ solution asymptotically locally AdS$_7/\mathbb{Z}_k$.}
\label{2ddefect}
\end{figure}

For our choice   $\beta_0=\frac{\alpha_0}{2}(n+1)$ we have, according to (\ref{alphaz}), 
\begin{equation}
\label{alphay}
\alpha(y)=\frac{\alpha_0}{2}y (n+1-y)\, , \qquad \dot{\alpha}(y)=\frac{\alpha_0}{2} (n+1-2y)\, .
\end{equation}
One can see from these expressions that $\alpha(y)$ takes its maximum value at $y=\frac{n+1}{2}$, and that it is symmetric under $y\leftrightarrow n+1-y$. We have for $y=j$, $\alpha(j)=\frac{\alpha_0}{2}\, j (n+1-j)$, and $\dot{\alpha}(j)=\frac{\alpha_0}{2}(n+1-2j)$. 
Using this we can now compute the D2 and D4 brane charges.
The D2-branes are stretched between NS5-branes located at $y=j, j+1$, for $j=1,\dots n$. Between them there are perpendicular D4-branes. Using (\ref{F6quiver}) and (\ref{F4quiver}) we then find, in the $[j,j+1]$ interval
\begin{equation}
Q_{\text{D}2}^{(j)}=\frac{1}{(2\pi)^5}\int_{I_\mu\times S^3/\mathbb{Z}_k\times S^2}{\hat F}_{(6)}= \frac{10\,k'}{k}\, j(n+1-j)\int e^{6U} d\mu
\end{equation}
and 
\begin{equation}
Q_{\text{D}4}^{(j)}=\frac{1}{(2\pi)^3}\int_{I_\mu\times S^3/\mathbb{Z}_k}{\hat F}_{(4)}= \frac{10\,k'}{k}(n+1-2j)\int e^{6U} d\mu
\end{equation}
where we have used expressions (\ref{chargedDWsol7d}) together with $\alpha_0=81\pi^2 Q_{\text{D}6}=81\pi^2 k'$.
The variation in the number of D4-branes from the $j$'th to the $(j+1)$'th interval is then  
\begin{equation}
\Delta Q_{\text{D}4}^{(j)}=Q_{\text{D}4}^{(j)}-Q_{\text{D}4}^{(j+1)}=\frac{20\,k'}{k}\int e^{6U} d\mu.
\end{equation}
As expected, the D2-D4 defect sees the infinity coming from the non-compactness of the $\mu$-direction. This is translated into large quantised charges for the D2 and D4 branes, the regime in which the $\mathrm{AdS}_3$ solutions can be trusted. We define $N\equiv \frac{10\, k'}{k}\int e^{6U} d\mu$. In terms of this new parameter the D2 and D4-brane charges read
\begin{equation}
\label{N2N4}
Q_{\text{D}2}^{(j)}=j(n+1-j)N\, , \qquad Q_{\text{D}4}^{(j)}=(n+1-2j)N\, , \qquad \Delta Q_{\text{D}4}^{(j)}=2N\, .
\end{equation}
Together with the charges coming from the D6-branes, $Q_{\text{D}6}=k'$, these quantised charges give rise to a non-anomalous 2d quiver CFT, that we have depicted in Figure \ref{2ddefect}, where we have denoted $P\equiv (n+1)/2$. 
This quiver is of the type recently discussed in \cite{Lozano:2019jza,Lozano:2019zvg}, whose main properties we have summarised in appendix \ref{summary2dCFT}. These quivers consist on gauge nodes associated to colour D2 and D6 branes to which flavour groups associated to D4 branes can be attached. The specific vector and matter fields that enter in the quivers are summarised in appendix \ref{summary2dCFT}, together with the anomaly cancelation conditions of the associated chiral 2d CFTs. In the quiver depicted in Figure \ref{2ddefect} the D2-branes contribute with the gauge nodes in the upper row. These couple to the gauge nodes associated to the D6-branes, in the lower row, through (0,4) hypermultiplets (the vertical lines) and (0,2) Fermi multiplets (the diagonal lines). In turn, the flavour groups  associated to the D4-branes couple to the later gauge nodes by means of  (0,2) flavour Fermi multiplets. These specific couplings of the vector and matter fields associated to the different branes finally render the 2d quiver CFT non-anomalous (see below). Note that in order to achieve this the gauge and flavour groups associated to the D2-D4 defect branes need to couple quite non-trivially to the gauge and flavour groups associated to the D6-branes of the mother 6d CFT, depicted in Figure  \ref{6dquiver}. One can see in particular that it is not possible to detach a 2d CFT built out from just the D2-D4 branes. In turn, the 6d quiver CFT depicted in Figure \ref{6dquiver} can be decoupled from the D2 and the D4 branes. These facts are fully consistent with our defect interpretation of the solution.  

Finally, we check that the quiver CFT satisfies the anomaly cancellation conditions for 2d ${\mathcal N}=(0,4)$ SCFTs, briefly summarised in appendix \ref{summary2dCFT}.
According to equation (\ref{anomaly}) we trivially have, for the $\text{SU}(Q_{\text{D}2}^{(j)})$ gauge groups, $2k'=k'+k'$.  Extra $k'$ flavour groups need to be attached to the $\text{SU}(Q_{\text{D}2}^{(1)})$ and $\text{SU}(Q_{\text{D}2}^{(n)})$ gauge groups, that are associated to the $k'$ D6-branes that terminate the space at $y=0, n+1$. In turn, for the $\text{SU}(k')$ gauge groups we can easily see that the anomaly cancelation condition
\begin{equation}
2Q_{\text{D}2}^{(j)}=Q_{\text{D}2}^{(j-1)}+Q_{\text{D}2}^{(j+1)}+\Delta Q_{\text{D}4}^{(j)}\, ,
\end{equation}
is satisfied for the charges in equation (\ref{N2N4}).


\vspace{0.3cm}

\noindent {\bf Central charge}:
\vspace{0.2cm}

\noindent At the conformal point the (right moving) central charge of a 2d ${\mathcal N}=(0,4)$ QFT  is related to the $U(1)_R$ current correlation function (see for example \cite{Putrov:2015jpa}), such that
\begin{equation}
c=6(n_{hyp}-n_{vec}),
\end{equation}
where $n_{hyp}$ is the number of $\mathcal{N}=(0,4)$ hypermultiplets and $n_{vec}$ the number of 
$\mathcal{N}=(0,4)$ vector multiplets of the theory in its UV description.

For the quiver depicted in Figure \ref{2ddefect} we have
\begin{equation}
n_{hyp}=\sum_{j=1}^n Q_{\text{D}2}^{(j)}Q_{\text{D}2}^{(j+1)}+(n+1) Q_{\text{D}6}^2+Q_{\text{D}6}\sum_{j=1}^{n+1} Q_{\text{D}2}^{(j)}
\end{equation}
and
\begin{equation}
n_{vec}=\sum_{j=1}^{n+1} \Bigl((Q_{\text{D}2}^{(j)})^2-1\Bigr)+(n+1)(Q_{\text{D}6}^2-1)\, .
\end{equation}
It is easy to check that for large quivers the contribution of the vector multiplets cancels the contributions of the $Q_{\text{D}2}^{(j)}Q_{\text{D}2}^{(j+1)}$ and $Q_{\text{D}6}^2$ bifundamentals, leaving, to leading order in $n$, 
\begin{equation}
\label{cfieldtheory}
c\sim 6\, Q_{\text{D}6} \,N \sum_{j=1}^{n+1} j (n+1-j)\sim Q_{\text{D}6}\, Q_{\text{NS}5}^3\, N=
\frac{1}{k} Q_{\text{D}6}^2 \, Q_{\text{NS}5}^3\, N'
\end{equation}
where we have used that $Q_{\text{NS}5}=n+1$ and redefined $N\equiv \frac{k'}{k} N'$. Therefore, the central charge diverges cubically with the number of nodes in the quiver, and quadratically with the number of D6-branes. Moreover, it diverges due to the non-compactness of the $\mu$-direction. This divergence is absorbed in the parameter $N'$. This second divergence is of interest physically, because it shows explicitly that the 2d quiver CFT per-se is ill-defined. This pathological behaviour of the central charge is cured in the UV, by the emergence of the deconstructed extra dimensions where the 6d CFT lives. 

This is supported by the behaviour of the holographic central charge.
Using expression (\ref{cdefinition}), whose derivation is summarised in appendix~\ref{newAdS3}, we find for the backgrounds defined by (\ref{metricAdS7X}), (\ref{dilatonAdS7X}),
\begin{equation}
c_{hol}=\frac{2^9 (-\ddot{\alpha})}{3^7 \pi^4 g^4 k}\int dy\, d\mu\, \alpha\, e^{4U+V}\, .
\end{equation}
This expression reproduces exactly the $\frac{1}{k}Q_{\text{D}6}^2Q_{\text{NS}5}^3$ behaviour in (\ref{cfieldtheory}), times an infinity that, as before, arises from the $\mu$-integration. Upon convenient regularisation both expressions can be found to agree. It would be interesting however to understand better the precise relation between the field theory and holographic central charges in ill-defined CFTs associated to defects. One could expect in particular that a non-trivial mixing between the holographic parameter and the energy scale could be at play.

\section{Surface defects in massive IIA} \label{massiveIIA}

The previous section was devoted to the study of the reduction of the 11d $\mathrm{AdS}_3$ solutions and brane set-up along the Taub-NUT direction of the KK'-monopoles, contained in the worldvolume of the M5'-branes. In this section we will be concerned with the reduction to Type IIA along the Taub-NUT direction $\chi$ of the second set of Kaluza-Klein monopoles, the KK-monopoles referred to in Table \ref{Table:branesinAd7}. As we already pointed out, this reduction destroys the $\mathrm{AdS}_7$ structure in 10d. This appears clear by looking at the near-horizon metric \eqref{brane_metric_M2M5KKM5_nh}, where the M-theory circle is taken within the 3-sphere $S^3/\mathbb{Z}_k$, which was part of $\mathrm{AdS}_7/\mathbb{Z}_k$. The solutions to Type IIA that arise in this reduction are the $\mathrm{AdS}_3\times S^2\times \mathrm{CY}_2$ solutions recently constructed in  \cite{Lozano:2019emq}, with $\mathrm{CY}_2=\mathbb{C}^2/\mathbb{Z}_{k'}$. This general class of solutions was constructed as solutions to massive IIA. Upon reduction from M-theory we recover the massless subclass.

In this section we show that these solutions can be given a defect interpretation when embedded in massive IIA. Therefore, we will be considering the general class of solutions constructed in  \cite{Lozano:2019emq}, with  $\mathrm{CY}_2=T^4$ locally. We will see that these solutions can be interpreted as associated to D4-KK'-D8 bound states on which smeared D2-NS5-D6 branes end. The D4-KK'-D8 brane system has as near-horizon geometry the $\mathrm{AdS}_6 \times S^4$ background of Brandhuber-Oz \cite{Brandhuber:1999np}, further orbifolded by $\mathbb{Z}_{k'}$, i.e. $\mathrm{AdS}_6\times S^4/\mathbb{Z}_{k'}$.

 Very much in analogy with the study carried out  in section \ref{7dDWchange}, we show that these solutions can be related to a 6d charged domain wall solution characterised by an $\mathrm{AdS}_3$ slicing and a 2-form gauge potential \cite{Dibitetto:2018iar}. This domain wall reproduces locally in its asymptotic regime the $\mrm{AdS}_6$ vacuum in \cite{Brandhuber:1999np} associated to D4-D8 branes, modded out by $\mathbb{Z}_{k'}$. In the opposite limit a singular behaviour describes D2-NS5-D6 brane sources that, intersecting the D4-D8-KK' system, create a defect, breaking the isometries of $\mrm{AdS}_6$ to those of $\mrm{AdS}_3$.

\subsection{The brane set-up}\label{branesmassiveIIA}

We start considering the well-known D4-D8 brane set-up of massive IIA string theory \cite{Brandhuber:1999np,Imamura:2001cr}, with D2-NS5-D6 branes ending on it \cite{Dibitetto:2018iar}. For the moment we will ignore the contribution of the KK'-monopoles, since they do not break any further supersymmetry and do not change substantially the properties of the background. We will include them later  by simply replacing the $\mathbb{R}^4$ transverse to the D4-branes by $\mathbb{R}^4/\mathbb{Z}_{k'}$, in the parametrisation $ds^2_{\mathbb{R}^4/\mathbb{Z}_{k'}}=d\rho^2+\rho^2\,ds^2_{\tilde S^3/\mathbb{Z}_{k'}}$. 

The D4-D8-D2-NS5-D6 branes set-up depicted in Table \ref{Table:branesinAd6} preserves 4 real supercharges. This is due to the presence of the D8-branes, which relate the charge distributions of the NS5 and D6 branes  \cite{Imamura:2001cr}. As we said, we are interested in a particular realisation of branes reproducing locally in the UV  the $\mrm{AdS}_6$ vacuum associated to the D4-D8 brane system.  
\begin{table}[h!]
\renewcommand{\arraystretch}{1}
\begin{center}
\scalebox{1}[1]{
\begin{tabular}{c||c c|c c c | c | c c c c}
branes & $t$ & $x^1$ & $r$ & $\theta^{1}$ & $\theta^{2}$ & $z$ & $\rho$ & $\varphi^1$ & $\varphi^2$ & $\phi$ \\
\hline \hline
$\mrm{D}8$ & $\times$ & $\times$ & $\times$ & $\times$ & $\times$ & $-$ & $\times$ & $\times$ & $\times$ & $\times$ \\
$\mrm{D}4$ & $\times$ & $\times$ & $\times$ & $\times$ & $\times$ & $-$ & $-$ & $-$ & $-$ & $-$ \\
$\mrm{D}6$ & $\times$ & $\times$ & $-$ & $-$ & $-$ & $\times$ & $\times$ & $\times$ & $\times$ & $\times$ \\
$\mrm{NS}5$ & $\times$ & $\times$ & $-$ & $-$ & $-$ & $-$ & $\times$ & $\times$ & $\times$ & $\times$ \\
$\mrm{D}2$ & $\times$ & $\times$ & $-$ & $-$ & $-$ & $\times$ & $-$ & $-$ & $-$ & $-$ \\
\end{tabular}
}
\end{center}
\caption{Brane picture underlying the intersection of D2-NS5-D6 branes ending on the D4-D8 brane system. The system is $\mrm{BPS}/8$.} \label{Table:branesinAd6}
\end{table}
To this end we consider the following 10d metric,
\begin{equation}
\label{brane_metric_D2D4NS5D6D8}
\begin{split}
d s_{10}^2&=H_{\mathrm{D}4}^{-1/2}\,H_{\mathrm{D}8}^{-1/2}\,\left[H_{\mathrm{D}6}^{-1/2}\,H_{\mathrm{D}2}^{-1/2}\,ds^2_{\mathbb{R}^{1,1}}+H_{\mathrm{D}6}^{1/2}\,H_{\mathrm{D}2}^{1/2} \,H_{\mathrm{NS}5}(dr^2+r^2d s^2_{S^2}) \right]\\
&+H_{\mathrm{D}4}^{1/2}\,H_{\mathrm{D}8}^{1/2}H_{\mathrm{D}6}^{-1/2}\,H_{\mathrm{D}2}^{-1/2} \,H_{\mathrm{NS}5}dz^2+H_{\mathrm{D}4}^{1/2}\,H_{\mathrm{D}8}^{-1/2}H_{\mathrm{D}6}^{-1/2}\,H_{\mathrm{D}2}^{1/2}(d\rho^2+\rho^2 ds^2_{\tilde S^3}) \, ,
\end{split}
\end{equation}
where we take the D2 and the NS5-branes smeared\footnote{The existence of this string background has been originally discussed in \cite{Dibitetto:2018iar}. Here we provide the explicit solution. We thank Niall Macpherson for a very useful discussion regarding this set-up and for pointing out the smearing of the NS5-branes.} over the space  transverse to the D4-branes, i.e. $H_{\mathrm{D}2}=H_{\mathrm{D}2}(r)$ and $H_{\mathrm{NS}5}=H_{\mathrm{NS}5}(r)$. Together with the metric \eqref{brane_metric_D2D4NS5D6D8}, we consider the following set of gauge potentials and dilaton,
\begin{equation}
\begin{split}\label{brane_potentials_D2D4NS5D6D8}
&C_{(3)}=H_{\mathrm{D}8}\,H_{\mathrm{D}2}^{-1}\,\text{vol}_{\mathbb{R}^{1,1}}\wedge dz\,,\\
&C_{(5)}=H_{\mathrm{D}6}\,H_{\mathrm{NS}5}\,H_{\mathrm{D}4}^{-1}\,r^2\,\text{vol}_{\mathbb{R}^{1,1}}\wedge dr \wedge \text{vol}_{S^2}\,,\\
&C_{(7)}=H_{\mathrm{D}4}\,H_{\mathrm{D}6}^{-1}\,\rho^3\,\text{vol}_{\mathbb{R}^{1,1}}\wedge dz\wedge d\rho \wedge \text{vol}_{\tilde S^3} \,,\\
&B_{(6)}=H_{\mathrm{D}8}\,H_{\mathrm{D}4}\,H_{\mathrm{NS}5}^{-1}\,\rho^3\,\text{vol}_{\mathbb{R}^{1,1}}\wedge  d\rho \wedge\text{vol}_{\tilde S^3}\,,\\\vspace{0.4cm}
&e^{\Phi}=H_{\mathrm{D}8}^{-5/4}\,H_{\mathrm{D}4}^{-1/4}\,H_{\mathrm{D}6}^{-3/4}\,H_{\mathrm{NS}5}^{1/2}\,H_{\mathrm{D}2}^{1/4}\,,
\end{split}
\end{equation}
with the $C_{(9)}$ potential for D8 branes defining the Romans mass as $F_{(0)}=m$.
One can then derive the fluxes\footnote{We use the conventions of \cite{Imamura:2001cr}.
} 
\begin{equation}
\begin{split}
&F_{(0)}=m\,,\\
 &F_{(2)}=H_{\mathrm{D}8}\,\partial_r\, H_{\mathrm{D}6}\,r^2\,\text{vol}_{S^2}\,, \\
 &H_{(3)}=\partial_r\,H_{\mathrm{NS}5}\, r^2\,\text{vol}_{S^2}\wedge dz\,,\\
 &F_{(4)}=H_{\mathrm{D}8}\,\partial_r\,H_{\mathrm{D}2}^{-1}\,\text{vol}_{\mathbb{R}^{1,1}}\wedge dr\wedge dz+H_{\mathrm{D}2}\,H_{\mathrm{NS}5}^{-1}\,\partial_z H_{\mathrm{D}4}\rho^3\,d\rho\wedge \text{vol}_{\tilde S^3}-H_{\mathrm{D}8}\partial_\rho H_{\mathrm{D}4}\rho^3\,dz \wedge \text{vol}_{\tilde S^3}
 \end{split}
\end{equation}
for which the Bianchi identities for $F_{(2)}$ and $H_{(3)}$ take the form
\begin{equation}\label{bianchiD2D4NS5D6D8}
 \partial_zH_{\mathrm{D}8}=m\,,\qquad H_{\mathrm{NS}5}=H_{\mathrm{D}6}=H_{\mathrm{D}2}\,,\qquad \nabla^2_{\mathbb{R}^3_r}\,H_{\mathrm{NS}5}=0\,.
\end{equation}
Imposing the relations \eqref{bianchiD2D4NS5D6D8}, the Bianchi identities for $F_{(4)}$ and the equations of motion collapse to the equation describing the D4-D8 system \cite{Imamura:2001cr},
\begin{equation}
\begin{split}\label{eomD4D8}
 &H_{\mathrm{D}8}\,\nabla_{T^4}^2\,H_{\mathrm{D}4}+\partial_{z}^2\,H_{\mathrm{D}4}=0\,.
 \end{split}
\end{equation}
We can finally write down a particular solution as 
\begin{equation}\label{solD2D4NS5D6D8}
  H_{\mathrm{NS}5}(r)=1+\frac{Q_{\mathrm{NS}5}}{r}\,,\qquad H_{\mathrm{D}6}(r)=1+\frac{Q_{\mathrm{D}6}}{r}\,,\qquad  H_{\mathrm{D}2}(r)=1+\frac{Q_{\mathrm{D}2}}{r}\,,
 \end{equation}
where $Q_{\mathrm{D}6}=Q_{\mathrm{D}2}=Q_{\mathrm{NS}5}$ for \eqref{bianchiD2D4NS5D6D8} to be satisfied.

Let us consider now the limit $r\rightarrow 0$. In this regime the 10d background \eqref{brane_metric_D2D4NS5D6D8} takes the form\footnote{We redefined the Minkowski coordinates as $(t,x^1)\rightarrow 2\,Q_{\mathrm{NS}5}^{3/2}\,(t,x^1)\,.$} 
\begin{equation}
\begin{split}\label{nhsolD2D4NS5D6D8}
 &ds_{10}^2=H_{\mathrm{D}4}^{-1/2}\,H_{\mathrm{D}8}^{-1/2}\,Q_{\mathrm{NS}5}^2\,\left(4\,ds^2_{{\scriptsize \mathrm{AdS}_3}}+ds^2_{S^2}  \right)+H_{\mathrm{D}4}^{1/2}\,H_{\mathrm{D}8}^{1/2}dz^2+H_{\mathrm{D}4}^{1/2}\,H_{\mathrm{D}8}^{-1/2}(d\rho^2+\rho^2 ds^2_{\tilde S^3})\,,\\
 &e^{\Phi}=H_{\mathrm{D}8}^{-5/4}\,H_{\mathrm{D}4}^{-1/4}\,,\\
 &F_{(0)}=m\,,\\
 &F_{(2)}=-Q_{\mathrm{NS}5}\,H_{\mathrm{D}8}\,\text{vol}_{S^2}\,, \\
 &H_{(3)}=-Q_{\mathrm{NS}5}\,dz\wedge\text{vol}_{S^2}\,,\\
 &F_{(4)}=8\,Q_{\mathrm{NS}5}^{2}\,H_{\mathrm{D}8}\text{vol}_{{\scriptsize \mathrm{AdS}_3}}\wedge dz+\partial_z H_{\mathrm{D}4}\rho^3\,d\rho\wedge \text{vol}_{\tilde S^3}-H_{\mathrm{D}8}\partial_\rho H_{\mathrm{D}4}\rho^3\,dz \wedge \text{vol}_{\tilde S^3}\,.
 \end{split}
\end{equation}
In this limit the supergravity solution describes a D4-D8 system wrapping an $\mrm{AdS}_3 \times S^2$ geometry. As shown in \cite{Lozano:2019emq}, when this system is put in this curved background D2-D4-NS5 branes need to be added in order to preserve supersymmetry. The number of supersymmetries is then reduced to $\mathcal{N}=(0,4)$. The $\mathrm{AdS}_3$ background in \eqref{nhsolD2D4NS5D6D8} is indeed included in the classification of $\ma N=(0,4)$ $\mrm{AdS}_3\times S^2\times \text{CY}_2\times I$ solutions  found in \cite{Lozano:2019emq}. In particular, it can be reproduced from the class I of $\mrm{AdS}_3$ solutions written in (3.1) of \cite{Lozano:2019emq} for the case of  $\text{CY}_2=T^4$, $u^\prime=0$ and $H_2=0$, after the redefinitions,
\begin{equation}
\begin{split}\label{matchingnh-AdS3sol}
& H_{\mathrm{D}8}=\frac{h_8}{2Q_{\mathrm{NS}5}}\,, \qquad H_{\mathrm{D}4}=\frac{2^5Q_{\mathrm{NS}5}^5}{u^2}\,h_4\qquad\text{and}\qquad z=\frac{\tilde{\rho}}{2Q_{\mathrm{NS}5}}\,, \qquad \rho=\frac{u^{1/2}}{2^{3/2}Q_{\mathrm{NS}5}^{3/2}}\,\tilde r\,.
 \end{split}
\end{equation}
As we mentioned, substituting the $\tilde S^3$ with the Lens space  $\tilde S^3/\mathbb{Z}_{k'}$ in \eqref{nhsolD2D4NS5D6D8} one gets the near-horizon regime including KK'-monopoles.

A similar D2-D4-NS5-D6-D8 brane intersection to the one considered in this section was studied in 
\cite{Dibitetto:2017klx}. This brane intersection was obtained as a generalisation of the massless solution of \cite{Boonstra:1998yu} to include D8-branes. In these set-ups D2 branes are completely localised in their transverse space, and the system finds an interpretation in terms of D2-D4 defect branes ending on NS5-D6-D8 branes. Consistently with this interpretation, it was shown in \cite{Dibitetto:2017klx} that the corresponding $\ma N=(0,4)$ $\mathrm{AdS}_3$ near-horizon geometry asymptotes locally to the $\mathrm{AdS}_7$ vacuum of massive IIA supergravity \cite{Apruzzi:2013yva}.

\subsection{Surface defects as 6d curved domain walls}
\label{6dDW}
In this section we show that the $\mrm{AdS}_3$ background \eqref{nhsolD2D4NS5D6D8} describes, in a particular limit, the $\mathrm{AdS}_6$ vacuum associated to the D4-D8 system. The idea is to describe the geometry \eqref{nhsolD2D4NS5D6D8} in terms of a 6d domain wall characterised by an $\mrm{AdS}_3$ slicing and an asymptotic behaviour locally reproducing the $\mrm{AdS}_6$ vacuum.
This solution was found in \cite{Dibitetto:2017klx} in the context of 6-dimensional $\ma N=(1,1)$ minimal gauged supergravity (see appendix \ref{6dsugra} for more details on the theory and its embedding in massive IIA). 

We consider the following 6d background
\begin{equation}
\begin{split}\label{6dAdS3}
& ds^2_6=e^{2U(\mu)}\left(4\,ds^2_{{\scriptsize \mrm{AdS}_3}}+ds^2_{S^2} \right)+e^{2V(\mu)}d\mu^2\,,\\
&\ma B_{(2)}=b(\mu)\,\text{vol}_{S^2}\,,\\
&X_6=X_6(\mu)\,.
\end{split}
\end{equation}
This background is described by the following set of BPS equations \cite{Dibitetto:2017klx},
\begin{equation}
 \begin{split}
  U^\prime= -2\,e^{V}\,f_6\,,\qquad X_6^\prime=2\,e^{V}\,X_6^2\,D_Xf_6\,,\qquad b^\prime= \frac{e^{U+V}}{X_6^2}\,,
  \label{chargedDW6d}
 \end{split}
\end{equation}
together with the duality constraint
\begin{equation}\label{chargedDW6d1}
 b=-\frac{e^{U}\,X_6}{m}\,,
\end{equation}
and the superpotential $f_6$ written in \eqref{6dsuperpotential}. This flow preserves 8 real supercharges (BPS/2 in 6d). In order to obtain an explicit solution of \eqref{chargedDW6d}, a parametrisation of the 6d geometry needs to be chosen. The simplest choice is given by
\begin{equation}
 e^{-V}=2\,X_6^2\,D_Xf_6\,.
\end{equation}
The system \eqref{chargedDW6d} can then be integrated out easily \cite{Dibitetto:2017klx}, to give
\begin{equation}
 \begin{split}
  e^{2U}= &\ 2^{-1/3}g^{-2/3}\,\left(\frac{\mu}{\mu^4-1}\right)^{2/3}\ , \qquad e^{2V}=8\,g^{-2}\, 
  \frac{\mu^4}{\left( \mu^4-1\right)^2}\ ,\\
   b=&\ -2^{4/3}\,3\,g^{-4/3}\,\frac{\mu^{4/3}}{(\mu^4-1)^{1/3}}\ ,\qquad \ X_6=\mu\ ,
   \label{chargedDWsol}
 \end{split}
\end{equation}
with $\mu$ running between 0 and 1 and $m= \frac{\sqrt{2}}{3}g$. 

One can see that for $\mu \rightarrow 1$ the 6d background is such that
\begin{equation}
 \begin{split}
  \ma {R}_{6}= -\frac{20}{3}\,g^2+\ma O (1-\mu)^{2/3}\,,\qquad X_6=&\ 1+\ma O (1-\mu)\ ,
  \label{UVchargedDW6d}
 \end{split}
\end{equation}
where $\mathcal{R}_{6}$ is the scalar curvature. These are the curvature and scalar fields reproducing the $\mrm{AdS}_6$ vacuum  \eqref{BOvacuum}. In turn, the 2-form gauge potential gives non-zero sub-leading contributions in this limit. This implies that the asymptotic geometry for $\mu \rightarrow 1$ is only locally $\mrm{AdS}_6$.  In the opposite limit $\mu\rightarrow 0$, the 6d background is manifestly singular. This is due to the presence of the D2-NS5-D6 brane sources.

Let us consider now the truncation ansatz of massive IIA supergravity \eqref{truncationansatz6d} and \eqref{10dfluxesto6d} for the above 6d background,
\begin{equation}
 \begin{split}\label{uplift6dDW}
  ds^2_{10}&=s^{-1/3}\,X_6^{-1/2}\,\Sigma_6^{1/2}\,e^{2U}\left(4\,ds^2_{{\scriptsize \mrm{AdS}_3}}+ds^2_{S^2} \right)+s^{-1/3}\,X_6^{-1/2}\,\Sigma_6^{1/2}e^{2V}d\mu^2\\
  &+2g^{-2}s^{-1/3}\Sigma_6^{1/2}\,X_6^{3/2}\,d\xi^2+2g^{-2}\,X_6^{-3/2}\,\Sigma_6^{-1/2}\,s^{-1/3}\,c^{2}\,ds^2_{\tilde S^3}\ ,\\
  F_{(4)}&=-\frac{4\sqrt 2}{3}\,g^{-3}\,s^{1/3}\,c^3\,\Sigma_6^{-2}\,U\,d\xi\,\wedge\,\text{vol}_{\tilde S^3}-8\sqrt{2}\,g^{-3}\,s^{4/3}\,c^4\,\Sigma_6^{-2}\,X_6^{-3}\,X_6^\prime\,d\mu\,\wedge\,\text{vol}_{\tilde S^3}\\
  &-8\,\sqrt2 \,g^{-1}\,s^{1/3}\,c\,X_6^4\,b^\prime\,e^{U-V}\,d\xi \wedge \text{vol}_{{\scriptsize \mrm{AdS}_3}}-8\, m\,s^{4/3}\,b\,X_6^{-2}\,e^{U+V}\,d\mu \wedge \text{vol}_{{\scriptsize \mrm{AdS}_3}}\,,\\
  F_{(2)}&=m\,s^{2/3}\,b\,\text{vol}_{S^2}\ ,\qquad H_{(3)}=s^{2/3}\,b^\prime\,d\mu \wedge \text{vol}_{S^2}+\frac{2}{3}\,s^{-1/3}\,c\,b\,d\xi \wedge \text{vol}_{S^2}\ ,\\
  e^{\Phi}&=s^{-5/6}\,\Sigma_6^{1/4}\,X_6^{-5/4}\ ,\qquad F_{(0)}=m\,,
 \end{split}
\end{equation}
with $c=\cos\xi$, $s=\sin \xi\,,\,\,\Sigma_6=X_6\,c^2+X_6^{-3}\,s^2$ and $U$ given by \eqref{10dfluxesto6d}. It is possible to show that the background \eqref{uplift6dDW} takes exactly the form of the near-horizon metric \eqref{nhsolD2D4NS5D6D8}. For this one needs to perform the change of coordinates
\begin{equation}\label{coord6dAdS6}
  z=\frac{3\,s^{2/3}\,e^U\,X_6}{\sqrt{2}\,g\, Q_{\mathrm{NS}5}}\,, \qquad \rho=\frac{\sqrt 2\,c\,e^{3U/2}}{g\,Q_{\mathrm{NS}5}^{3/2}\,X_6^{1/2}}\,,
\end{equation}
and use the 6d BPS equations \eqref{chargedDW6d}, \eqref{chargedDW6d1}. We can thus express the warp factors describing the D4 and D8 branes in \eqref{nhsolD2D4NS5D6D8} in terms of the 6d domain wall realising the defect, as 
\begin{equation} \label{restH8H4}
 H_{\mathrm{D}8}=\frac{s^{2/3}\,e^U\,X_6}{Q_{\mathrm{NS}5}}\,,\qquad H_{\mathrm{D}4}=\frac{Q_{\mathrm{NS}5}^5\,e^{-5U}}{\Sigma_6}\,.
\end{equation}
One can check that these expressions satisfy the equations of motion for $H_{\mathrm{D}4}$ and $H_{\mathrm{D}8}$ written in \eqref{eomD4D8}.

We have thus shown that the $\mrm{AdS}_3$ background \eqref{nhsolD2D4NS5D6D8}, describing the near-horizon limit of D2-NS5-D6 branes ending on the D4-D8 brane system, reproduces locally the $\mrm{AdS}_6$ vacuum of \cite{Brandhuber:1999np}, for  $H_{\mathrm{D}8}$, $H_{\mathrm{D}4}$ given by (\ref{restH8H4}). This vacuum geometry comes out thanks to a non-linear mixing of the $(z,\rho)$ coordinates, that relates the near-horizon geometry to a 6d domain wall admitting $\mrm{AdS}_6$ in its asymptotics. The presence of the 2-form does not allow however to globally recover the vacuum in this limit. This is seen explicitly at the level of the uplift \eqref{uplift6dDW}, where one notes that the $F_{(2)}$ and $H_{(3)}$ fluxes break  the isometries of the D4-D8 vacuum. This is the manifestation of the D2-NS5-D6 defect, that underlies as well the singular behaviour of the 6d domain wall in its IR regime.

\section{Line defects in massive IIA} \label{line-defects}

Very much in analogy with our previous analysis, we show in this section that the AdS$_2$ solutions to massive IIA supergravity recently constructed in \cite{Lozano:2020bxo} can be given a  line defect CFT interpretation within the Brandhuber-Oz system. 
The solutions studied in \cite{Lozano:2020bxo} were obtained through double analytical continuation from the $\mathrm{AdS}_3\times S^2 \times \text{CY}_2\times I$ backgrounds constructed in \cite{Lozano:2019emq}. We showed in the previous section that a subset of these backgrounds with $\text{CY}_2=T^4$ reproduces locally the $\mrm{AdS}_6$ vacuum of \cite{Brandhuber:1999np}, thus allowing for a surface defect interpretation. In this section we show that the solutions in \cite{Lozano:2020bxo} with $\text{CY}_2=T^4$ can be given a similar defect interpretation within the D4-D8 brane system, this time as line defects.

Following the same spirit of the previous sections, a brane solution related to the $\mathrm{AdS_2}$ geometries mentioned above was worked out in \cite{Dibitetto:2018gtk}. This brane solution describes  
a D0-F1-D4' bound state ending on D4-D8 branes, as depicted in Figure \ref{Table:D0F1D4D4D8}.
\begin{table}[http!]
\renewcommand{\arraystretch}{1}
\begin{center}
\scalebox{1}[1]{
\begin{tabular}{c||c|c c c c|c||c c c c}
branes & $t$ & $r$ & $\theta^{1}$ & $\theta^{2}$ & $\theta^{3}$  & $z$ & $\rho$ & $\varphi^{1}$ & $\varphi^{2}$ & $\varphi^{3}$\\
\hline \hline
D8 & $\times$ & $\times$ & $\times$ & $\times$ & $\times$ & $-$ & $\times$ & $\times$ & $\times$ & $\times$  \\
D4 & $\times$ & $\times$ & $\times$ & $\times$ & $\times$ & $-$ & $-$ & $-$ & $-$ & $-$\\
D0 & $\times$ & $-$ & $-$ & $-$ & $-$ & $-$ & $-$ & $-$ & $-$ & $-$ \\
F1 & $\times$ & $-$ & $-$ & $-$ & $-$ & $\times$ & $-$ & $-$ & $-$ & $-$ \\
D4' & $\times$ & $-$ & $-$ & $-$ & $-$ & $-$ & $\times$ & $\times$ & $\times$ & $\times$
\end{tabular}
}
\end{center}
\caption{The brane picture of D0-F1-D4' branes ending on the D4-D8 system \cite{Dibitetto:2018gtk}. The intersection is BPS/8.} \label{Table:D0F1D4D4D8}
\end{table}
As in the calculation in section \ref{branesmassiveIIA}, allowing the D4-branes to be completely localised in their transverse space, it is possible to recover a near-horizon geometry describing a D4-D8 system wrapping an $\mrm{AdS}_3\times S^2$ geometry, to which D0-F1-D4' branes need to be added to preserve supersymmetry \cite{Dibitetto:2018gtk}. The near-horizon reads
\begin{equation}\label{D8D4D0F1D4'-nh}
ds_{10}^{2}  =  H_{\textrm{D}4}^{-1/2}H_{\textrm{D}8}^{-1/2}\left[Q_{1} \left(ds_{\textrm{AdS}_{2}}^{2}+4ds_{S^{3}}^{2}\right)+H_{\textrm{D}4}H_{\textrm{D}8}dz^{2}+
H_{\textrm{D}4}\,\left (d\rho^{2}+\rho^{2}\,ds_{\tilde{S}^{3}}^{2}\right)\right] \, ,
\end{equation}
with $Q_1$ a parameter related to the defect charges of D0-F1-D4' branes.
One can check that this background is included in the classification found in (5.1) of \cite{Lozano:2020bxo}, for $\text{CY}_2=T^4$ locally and $u^\prime=0$, after the redefinitions given by (\ref{matchingnh-AdS3sol}). 

Further, the previous brane intersection was linked in \cite{Dibitetto:2018gtk} to a 6d charged domain wall characterised by an $\mrm{AdS}_2$ slicing flowing asymptotically to the  $\mrm{AdS}_6$ vacuum of 6d Romans supergravity (see appendix~\ref{6dsugra}). This domain wall is of the form
\begin{equation}
\begin{split}\label{6dAdS2}
& ds^2_6=e^{2U(\mu)}\left(ds^2_{{\scriptsize \mrm{AdS}_2}}+4ds^2_{S^3} \right)+e^{2V(\mu)}d\mu^2\,,\\
&\ma B_{(2)}=b(\mu)\,\text{vol}_{{\scriptsize \mrm{AdS}_2}}\,,\\
&X_6=X_6(\mu)\,,
\end{split}
\end{equation}
and, consistently with the whole picture, can be obtained through double analytical continuation from the domain wall solution in \eqref{6dAdS3}. The BPS equations for this background preserve 8 real supercharges and take the same form of \eqref{chargedDW6d} and \eqref{chargedDW6d1}. In analogy with the $\mrm{AdS}_3$ analysis, the 6d solution \eqref{6dAdS2} reproduces locally in the limit $\mu \rightarrow 1$ the geometry of the $\mrm{AdS}_6$ vacuum, together with a singularity in the $\mu \rightarrow 0$ limit. Using the uplift formulas  to massive IIA given in \eqref{truncationansatz6d} one can check that the resulting domain wall solution in 10d is related to the near horizon geometry \eqref{D8D4D0F1D4'-nh} through the change of coordinates \cite{Dibitetto:2018gtk}
\begin{equation}
  z=\frac{3\,s^{2/3}\,e^U\,X_6}{\sqrt{2}\,g\, Q_1^{1/2}}\,, \qquad \rho=\frac{\sqrt 2\,c\,e^{3U/2}}{g\,Q_1^{3/4}\,X_6^{1/2}}\,,
\label{coordchangeAdS2}
\end{equation}
and the requirements for the $H_{\textrm{D}4}$ and $H_{\textrm{D}8}$ functions
\begin{equation} \label{H8H4AdS2}
 H_{\mathrm{D}8}=\frac{s^{2/3}\,e^U\,X_6}{Q_1^{1/2}}\,,\qquad H_{\mathrm{D}4}=\frac{Q_{1}^{5/2}\,e^{-5U}}{\Sigma_6}\,.
\end{equation}
These conditions are analogous to~\eqref{coord6dAdS6}-\eqref{restH8H4} for $\mrm{AdS}_3$, which is obviously related  to the fact that the $\mathrm{AdS}_2$ solutions and the $\mathrm{AdS}_3$ backgrounds discussed in the previous section are related by double analytical continuation.
In this case the solution is interpreted as a D0-F1-D4' line defect within the 5d Sp(N) fixed point theory.

\section{Conclusions}\label{conclusions}

In this paper we have obtained explicit brane intersections underlying different classes of AdS$_3$ solutions to Type IIA supergravity with $\mathcal{N}=(0,4)$ supersymmetries recently constructed in the literature. Furthermore, we have related these solutions to Janus-type domain wall backgrounds admitting asymptotic regions  described locally by higher dimensional AdS vacua. This has allowed us to provide a surface defect CFT interpretation for the AdS$_3$ solutions, where the mother CFT is the holographic dual of the higher dimensional AdS vacuum. 

We have analysed two classes of AdS$_3$ solutions with $\mathcal{N}=(0,4)$ supersymmetries. The first one is the class of AdS$_3\times S^3/\mathbb{Z}_k\times {\tilde S}^3\times \Sigma_2$ solutions to M-theory constructed in \cite{Lozano:2020bxo}, further orbifolded by $\mathbb{Z}_{k'}$.  These solutions are associated to M2-M5-M5' brane intersections, with the 5-branes placed in ALE singularities. We have found that a subclass of these solutions asymptote locally to the AdS$_7/\mathbb{Z}_k$ vacuum of 11d supergravity. This has allowed us to give a defect interpretation of these solutions in terms of M2-M5 branes (on an ALE singularity) embedded in M5'-branes on ALE singularities, realising a 6d (1,0) CFT. Upon reduction, we have found a new class of AdS$_3\times S^3/\mathbb{Z}_k\times S^2\times \Sigma_2$ solutions to Type IIA with $\mathcal{N}=(0,4)$ supersymmetries. We have found the right parametrisation that allows to interpret these solutions as 
holographic duals to surface defect CFTs. These originate from D2-D4 branes ending on the D6-NS5-KK brane intersection dual to the AdS$_7/\mathbb{Z}_k$ vacuum of massless IIA supergravity. We have presented an explicit 2d (0,4) quiver CFT that realises the D2-D4 defect CFT. In this quiver it is clear that the D2-D4 defect needs the D6-NS5-KK branes of the mother CFT in order to exist as a 2d CFT. Instead, from the 2d CFT the 6d mother CFT dual to the D6-NS5-KK intersection can be obtained in a certain decoupling limit. Finally, 
we have extended the previous class of solutions onto a more general class, obtained upon reduction of the AdS$_3\times S^3/\mathbb{Z}_k\times T^4\times I$ solutions to M-theory constructed in  \cite{Lozano:2020bxo}, further modded by $\mathbb{Z}_{k'}$. An interesting open problem is to find global completions of this more general class of solutions, which do not seem to asymptote locally to a higher dimensional AdS space. Work is in progress \cite{FLP} that shows that they can be completed in terms of globally well-defined AdS$_3$ solutions related upon a chain of T-S-T dualities to the AdS$_3\times S^2\times T^4\times I$ solutions recently constructed in \cite{Lozano:2019emq}. 

The second class of AdS$_3$ solutions with $\mathcal{N}=(0,4)$ supersymmetries that we have studied is the general classification of $\mathrm{AdS}_3\times S^2\times \mathrm{CY}_2\times I$ solutions to massive IIA supergravity constructed in \cite{Lozano:2019emq}, with $\mathrm{CY}_2=T^4$. We have provided the associated full brane solution and shown that it can be related to a 6d domain wall solution that reproduces asymptotically locally the AdS$_6$ vacuum of massive IIA supergravity. This has allowed us to interpret the solutions as holographic duals to surface defect CFTs originating from D2-NS5-D6 branes ending on D4-D8 bound states. It is likely that explicit quivers realising this can be constructed using the Type IIB description of the Sp(N) theory. This is currently being investigated \cite{FLP}.

Finally, and in full analogy with the previous analysis, we have discussed from the point of view of conformal defects a subclass of the $\mathrm{AdS}_2\times S^3\times T^4 \times I$ solutions with 4 supercharges recently constructed in \cite{Lozano:2020bxo}. Putting together previous results in the literature, that provided the full brane solution and linked it to a 6d domain wall reproducing asymptotically locally AdS$_6$, we have given an interpretation to these solutions as line defect CFTs originating from D0-F1-D4' branes embedded in the Brandhuber-Oz brane set-up. It would be interesting to find explicit realisations, possibly in terms of 1d ADHM-like  quantum mechanics as the ones described in \cite{Tong:2014cha,Kim:2016qqs}.

\section*{Acknowledgements}

We would like to thank Giuseppe Dibitetto, Niall Macpherson, Carlos Nunez and Anayeli Ramirez for very useful discussions. FF would like to thank the HEP Theory Group of the Universidad de Oviedo for its kind hospitality. The authors are partially supported by the Spanish government grant PGC2018-096894-B-100 and by the Principado de Asturias through the grant FC-GRUPIN-IDI/2018/000174.

\appendix

\section{M-theory origin of 7d $\ma N=1$ supergravity}\label{7dsugra}

In this appendix we recall the M-theory embedding of minimal $\ma N=1$ gauged supergravity in 7d \cite{Lu:1999bc}. The 7d theory preserves 16 supercharges and only the supergravity multiplet is retained by the truncation from 11d. All the oscillations around the $\mathrm{AdS}_7$ vacuum are thus encoded in the 7d gravitational field, a real scalar $X_7$, a 3-form gauge potential $\ma B_{(3)}$ and three $\mrm{SU(2)}$ vector fields $\ma A_7^i$ \cite{Townsend:1983kk}. We consider a further truncation of the theory in which all the vector fields are vanishing. This truncation ansatz has been worked out in \cite{Lu:1999bc}. It is characterised by an 11d metric of the following form
\begin{equation}
 \begin{split}\label{truncationansatz7d}
  &ds^2_{11}=\Sigma_7^{1/3}\,ds^2_7+2g^{-2}\,\Sigma_7^{-2/3}\,ds^2_{4}\,,\\
  &ds_{4}^2=X_7^3\,\Sigma_7\,d\xi^2+X_7^{-1}\,c^{2}\,ds^2_{S^3}\,,
 \end{split}
\end{equation}
where $\Sigma_7=X_7\,c^2+X_7^{-4}\,s^2$ with $c=\cos\xi$ and $s=\sin \xi$. The 11d 4-flux takes the form 
\begin{equation}
\begin{split}\label{truncationansatz7dfluxes}
  G_{(4)}&=-\frac{4}{\sqrt 2}\,g^{-3}\,c^{3}\,\Sigma_7^{-2}\,W\,d\xi\,\wedge\,\text{vol}_{S^3}-\frac{20}{\sqrt{2}}\,g^{-3}\,\Sigma_7^{-2}\,X_7^{-4}\,s\,c^4\,dX_7\,\wedge\,\text{vol}_{S^3}\\
  &+s\,\ma F_{(4)}+\sqrt{2}\,g^{-1}c\,X_7^{4}\,\star_{\,7}\ma F_{(4)}\wedge d\xi,\\
 \end{split}
\end{equation}
where $W=X_7^{-8}\,s^2-2X_7^2\,c^2+3\,X_7^{-3}\,c^2-4\,X_7^{-3}$ and $\ma F_{(4)}=d \ma B_{(3)}$. As it has been pointed out in \cite{Lu:1999bc}, in order to describe the right number of degrees of freedom, the 3-form $\ma B_{(3)}$ has to satisfy an ``odd-dimensional self-duality condition"
\begin{equation}
 X_7^4\,\star_{\,7}\ma F_{(4)}=-2h\,\ma B_{(3)}\,,\label{odddimselfdual}
\end{equation}
with $h=\frac{g}{2\sqrt 2}$ fixed by the truncation. The isometry group of the resulting 7d theory is given by $\mathbb{R}^+\times \mathrm{SO}(3)$ and there are two types of gaugings. One is described by the parameter $g$ and corresponds to the gauging of the R-symmetry $\mathrm{SU}(2)_R$, and the other is a St\"uckelberg deformation of $\ma B_{(3)}$ described by $h$. The general form of the superpotential is given by
\begin{equation}
\label{7dsuperpotential}
f_7(h,g,X_7) = \frac12 \left(h \, X_7^{-4} + \sqrt{2} \, g \, X_7 \right) \,,
\end{equation}
where the two gauging parameters are linked by the truncation through the algebraic relation $h=\frac{g}{2\sqrt 2}$. The Lagrangian is given by
\begin{equation}
\begin{split}
\label{7dlagrangian}
 \ma L_7&=  R_{\,7}  -5 \, X_7^{-2}\,\star_{\,7} dX_7 \wedge dX_7 - \frac12\,X_7^4\,\star_{\,7} \ma F_{(4)} \wedge \ma F_{(4)}- h \, \ma F_{(4)}\wedge\ma B_{(3)} - V_7
 \end{split}
\end{equation}
and the scalar potential by
\begin{equation}\label{7dpotential}
 V_7 = \frac{4}{5}\,X_7^{2}\,\left(D_{X}f_7\right)^{2} -\frac{24}{5}\,f_7^{2}\,.
\end{equation}
The theory \eqref{7dlagrangian} has a $\ma N=1$ $\mathrm{AdS}_7$ vacuum at $X_7=1$ and vanishing gauge potentials. In this case the internal 4d manifold of \eqref{truncationansatz7d} becomes a round 4-sphere. Since the theory \eqref{7dlagrangian} can be embedded into the maximally supersymmetric supergravity in seven dimensions, we can link this 7d vacuum to the $\mrm{AdS}_7\times S^4$ Freund-Rubin vacuum of M5-branes. In this particular case the 4-flux takes the form 
\begin{equation}\label{AdS7vacuum}
 G_{(4)}=-\frac{12}{\sqrt 2}\,g^{-3}\,c^{3}\,d\xi\,\wedge\,\text{vol}_{S^3}\,.
\end{equation}
On the contrary, the case considered in section \ref{7dDWchange} is that of the half-supersymmetric $\mathrm{AdS}_7$ vacuum of M-theory arising from M5-branes on an A-type singularity (associated to NS5-D6 branes in 10d). In this case the vacuum of \eqref{7dpotential} has to be interpreted in 11d as a half-maximal vacuum of the minimal $\ma N=1$ theory. 

\section{An extended class of AdS$_3\times S^3/\mathbb{Z}_k\times S^2$ solutions to Type IIA} \label{newAdS3}

In this appendix we extend the new class of AdS$_3\times S^3/\mathbb{Z}_k\times S^2$ solutions to Type IIA with ${\mathcal N}=(0,4)$ supersymmetries found in section 3.1. Our starting point is the general class of AdS$_3\times S^3/\mathbb{Z}_k \times \mathrm{CY}_2\times I$ solutions to M-theory constructed in \cite{Lozano:2020bxo}, with 
$\mathrm{CY}_2=T^4$ locally. The new solutions are obtained reducing on the Hopf-fibre of the 3-sphere contained locally in the $T^4$. This reduction preserves all the supersymmetries and generalises our solutions given by~\eqref{brane_metric_D2D4KKNS5D6_nh}. Prior to this reduction we extend the solutions in 
\cite{Lozano:2020bxo} by modding the 3-sphere by $\mathbb{Z}_{k'}$. 
This introduces $k'$ KK-monopoles that give rise to $k'$ D6-branes upon reduction. $k'$ can then be taken to be sufficiently large as in the IIA supergravity limit. 

The most general solutions in \cite{Lozano:2020bxo} are AdS$_3\times S^2\times \mathrm{CY}_2$ fibrations over two intervals (see appendix B therein). They take the form
\begin{equation}
\begin{split}
\label{m-theory-general-ClassI}
d s_{11}^2 &= \Delta \left( \frac{u}{\sqrt{h_4 h_8}} d s^2_{\text{AdS}_3} + \sqrt{\frac{h_4}{h_8}} d s^2_{\text{CY}_2} +\frac{\sqrt{{h_4 h_8}}}{u} d z^2 \right) + \frac{h_8^2}{4\Delta^2} \left( ds^2_{S^2} +\mathrm{D}\tilde{\chi}^2 \right)\, , \\
G_{(4)} &= - \left( d \left(\frac{u u^\prime}{2 h_4} \right) + 2 h_8 d z \right) \wedge \text{vol}_{\text{AdS}_3}  - \partial_{z} h_4 {\text{vol}}_{\text{CY}_2} - \frac{u u^\prime}{2 (h_4 h_8 + u'^2)} H_2 \wedge\text{vol}_{S^2} \\
&{\; \; \; \; }- \frac{h_8}{u} \star_4 d_4 h_4 \wedge d z+\frac{h_8}{2}\,\left[ \frac{1}{2} d \left( -z + \frac{u u^\prime}{4 h_4 h_8 + u'^2} \right) \wedge {\text{vol}}_{S^2} + \frac{1}{h_8} d z \wedge H_2  \right] \wedge \text{D}\tilde{\chi} \, ,
\end{split}
\end{equation}
where $H_2= -d\cal{A}$, $\text{D}\tilde{\chi}=d\tilde{\chi}+\tilde{\cal{A}}+\omega$, 
$d\omega={\text{vol}}_{S^2}$ and 
\begin{equation}
\Delta =\frac{h_8^{1/2}(4h_4h_8+u'^2)^{1/3}}{2^{2/3}h_4^{1/6}u^{1/3}}\, .
\end{equation}
In these solutions $h_8$ is a constant, $h_4$ has support on $(z, \text{CY}_2)$, $u$ is a function of $z$ and $H_2$ has support on the $\text{CY}_2$. Note that we have renamed $\rho$ and ${\tilde \psi}$ as in \cite{Lozano:2020bxo} by $z$ and ${\tilde \chi}$, respectively, to connect with our notation in section 3 (see below). The quantities
${\tilde \chi}$ and ${\tilde{\cal A}}$ are defined as $\tilde{\chi}= \frac{2}{h_8} \chi$ and $\tilde{\cal{A}} = \frac{2}{h_8}\cal{A}$. 
In the most general case the connection $\tilde{\cal{A}}+\omega$ makes the fibre over the $S^2$ and the $\mathrm{CY}_2$ non trivial.
Here we restrict to the case ${\cal A}=0$ and $\mathrm{CY}_2=T^4$. In this case the solutions simplify to
\begin{eqnarray}
\label{M-theory}
d s_{11}^2&=&\Delta\left(\frac{u}{\sqrt{h_4 h_8}} ds_{\text{AdS}_3}^2+\sqrt{\frac{h_4}{h_8}} \text{d}s_{T^4}^2+\frac{\sqrt{h_4 h_8}}{u} d z^2
\right)+\frac{h_8^2}{\Delta^2} d s^2_{S^3/\mathbb{Z}_k}\label{Mmetric} \, ,\\
G_{(4)}&=&-d\left(\frac{uu^\prime}{2 h_4}+2 h_8\,z\right)\wedge \text{vol}_{\text{AdS}_3}+2h_8\; d\left(-z+\frac{u u^\prime}{4h_4h_8+u'^2}\right)\wedge {\text{vol}}_{S^3/\mathbb{Z}_k}\nonumber \label{M4flux} \\
&&-\partial_{z}h_4\;{\text{vol}}_{T^4} - \frac{h_8}{u} \star_4 d_4 h_4 \wedge d z\, , 
\end{eqnarray}
where $k=h_8$ and $d s^2_{S^3/\mathbb{Z}_k}$ is written as in \eqref{orbifoldS3}.
Supersymmetry holds when
\begin{equation} 
\label{u}
u^{\prime\prime}=0\, ,
\end{equation}
while the Bianchi identities of the fluxes impose that
\begin{equation}
\label{Bianchi}
\frac{h_8}{u}\nabla^2_{T^4}h_4+\partial_z^2h_4=0.
\end{equation}
The symmetries $\text{SL}(2, \mathbb{R}) \times \text{SL}(2, \mathbb{R})$ and $\text{SU}(2)$ are realised geometrically on the AdS$_3$ and the quotiented 3-sphere, respectively. 

In order to extend the class of solutions given by (\ref{brane_metric_D2D4KKNS5D6_nh}) we take the local $T^4$ in~\eqref{M-theory}-\eqref{M4flux} to actually be $\mathbb{R}^4$ and mod it by $\mathbb{Z}_{k'}$, such that $ds^2_{\mathbb{R}^4/\mathbb{Z}_{k'}}=d\rho^2 + \rho^2\, ds^2_{{\tilde S}^3/\mathbb{Z}_{k'}}$ with $ds^2_{{\tilde S}^3/\mathbb{Z}_{k'}}$ defined in \eqref{orbifoldS3}.
The 4-flux term defined on the $\mathrm{CY}_2$ takes the form $ \star_4 d_4 h_4 =\rho^3 \,\partial_\rho h_4 {\rm vol}_{{\tilde S}^3/\mathbb{Z}_{k'}}$ and the equation (\ref{Bianchi}) imposed by the Bianchi identities becomes
\begin{equation}
\label{Bianchi2}
\frac{h_8}{u} \Bigl(\partial^2_\rho h_4+\frac{3}{\rho}\, \partial_\rho h_4\Bigr)+\partial^2_z h_4=0.
\end{equation}
Reducing now along the 
$S^1/\mathbb{Z}_{k'}\subset {\tilde S}^3/\mathbb{Z}_{k'}$ we obtain 
\begin{align}
ds^2_{10} &= \rho \left[{\tilde \Delta}  \left( \frac{u}{\sqrt{h_4 h_8}}ds^2_{\text{AdS}_3}+\frac{\sqrt{h_4 h_8}}{u}dz^2+
\sqrt{\frac{h_4}{h_8}}(d\rho^2 + \frac{\rho^2}{4}\, ds^2_{{\tilde S}^2})\right)+\frac{h_8^{3/2}h_4^{1/2}}{{\tilde \Delta}\,k'^2}ds^2_{S^3/\mathbb{Z}_k}\right] , \label{metricIIA}\\
e^{2\Phi} &= \frac{{\tilde \Delta}}{k'^2}\sqrt{\frac{h_4}{h_8}}\,\rho^3\, ,\qquad 
{\tilde \Delta}=\frac{1}{2k'}\sqrt{\frac{h_8(4h_4h_8+u'^2)}{u}}\,,\\
H_{(3)} &= -\frac{\rho^3}{4k'}\Bigl( \frac{h_8}{u}\partial_\rho h_4 dz-\partial_z h_4 d\rho\Bigr)\wedge {\rm vol}_{{\tilde S}^2}\,, \\
F_{(2)} &= \frac{k'}{2}{\rm vol}_{{\tilde S}^2}\,,\\
F_{(4)} &= -d\left(\frac{u u'}{2 h_4} + 2 h_8\,z\right) \wedge {\rm vol}_{\text{AdS}_3} + 2 h_8 \, d\left(-z + \frac{u u'}{4 h_4 h_8 + u'^2}\right) \wedge {\rm vol}_{S^3/\mathbb{Z}_k} \label{F4IIA} \, .
\end{align}
This extends the new class of 10d backgrounds with $\mathcal{N}=(0,4)$ supersymmetries presented in section 3.1 to include a new function $u(z)$, satisfying (\ref{u}). Indeed, one can check that the near horizon geometry  \eqref{brane_metric_D2D4KKNS5D6_nh} is obtained in the particular case $u^\prime=0$, with the redefinitions
\begin{equation}
Q_{\mathrm{KK}} = h_8 \,,  \qquad  Q_{\mathrm{D}6} = k' \,,  \qquad  H_{\mathrm{NS}5} = \frac{2^6 Q_{\mathrm{D}4}^3 h_8^2}{u^2} \, h_4  \qquad  \text{and}  \qquad  H_{\mathrm{D}6} = \frac{Q_{\mathrm{D}6}}{\rho} \,,
\end{equation}
where we rescaled the coordinates in~\eqref{brane_metric_D2D4KKNS5D6_nh} as
\begin{equation}
z \to \frac{z}{4 Q_{\mathrm{D}4}} \,,  \qquad  \rho \to \frac{u\,\rho^2}{2^5 Q_{\mathrm{D}6} Q_{\mathrm{D}4}^2 h_8} \,.
\end{equation}

As previously mentioned, doing the change of coordinates $\rho\rightarrow \rho^{1/2}$, one can see that there are $k'$ D6-branes seated at $ \rho=0$,
\begin{align}
ds^2_{10} &= \rho^{1/2} \left[ {\tilde \Delta}\left( \frac{u}{\sqrt{h_4 h_8}}ds^2_{\text{AdS}_3}+\frac{\sqrt{h_4 h_8}}{u}dz^2\right)+\frac{h_8^{3/2}h_4^{1/2}}{{\tilde \Delta}\,k'^2}ds^2_{S^3/\mathbb{Z}_k}\right]+\frac{{\tilde \Delta}}{4\rho^{1/2}}\sqrt{\frac{h_4}{h_8}}\left(d\rho^2+\rho^2 ds^2_{S^2}\right)\,, \nonumber \\
e^{2\Phi} &= \frac{{\tilde \Delta}}{k'^2}\sqrt{\frac{h_4}{h_8}}\, \rho^{3/2} \,,  \qquad  F_{(2)} = \frac{k'}{2}{\rm vol}_{S^2} \,.
\end{align}

Finally, we compute for completeness the holographic central charge of the new class of solutions.

\vspace{2mm}
\noindent \textbf{Holographic central charge:}
We recall that for a generic dilaton and background of the form
\begin{equation}
\label{holocc1}
ds^2=a(r,\vec{\theta})(dx_{\mathbb{R}^{1,d}}^2+b(r)dr^2)+g_{ij}(r,\vec{\theta})d\theta^i d\theta^j,\qquad \Phi(r,\vec{\theta}),
\end{equation}
the central charge can be computed\footnote{The factor of ``3" in (\ref{cdefinition}) is introduced as a normalisation, in order to match the standard result in \cite{Brown:1986nw}.} as \cite{Klebanov:2007ws,Macpherson:2014eza,Bea:2015fja}
\begin{equation}
\label{cdefinition}
c_{hol}=3\times \frac{d^d}{G_N}\frac{b(r)^{d/2}({\hat H})^{\frac{2d+1}{2}}}{({\hat H}')^d}\, ,
\end{equation}
where
\begin{equation}
{\hat H}=\Bigl(\int d{\vec \theta}\sqrt{e^{-4\Phi}{\rm det}[g_{ij}]a(r,\vec{\theta})^d}\Bigr)^2.
\end{equation}
Using Poincar\'e coordinates for AdS$_3$ we have for our solutions \eqref{metricIIA}-\eqref{F4IIA},
\begin{equation}
a(r,\vec{\theta})=\frac{\rho\,{\tilde \Delta}\,u}{\sqrt{h_4 h_8}} r^2, \qquad b(r)=\frac{1}{r^4},\qquad d=1\, ,
\end{equation}
and, finally,
\begin{equation}
\label{cfinal}
c_{hol}=\frac{3}{8G_N}\frac{k^2}{k'} {\rm Vol}_{S^2} {\rm Vol}_{S^3/\mathbb{Z}_k}\int h_4 \, \rho^3 \, d\rho\, dz=
\frac{3}{8\pi^3}\frac{k}{k'}\int h_4 \, \rho^3\, d\rho \, dz,
\end{equation}
where we have used that $G_N=8\pi^6$ and $h_8=k$. 

In order to obtain a finite result for the central charge using this expression we need a well-defined UV completion for our solutions. In the main body of the paper we have studied an interesting completion of a subclass of our solutions by which they flow in the UV into a higher dimensional $\mathrm{AdS}_7$ geometry, and can thus be  interpreted as surface CFTs within the higher dimensional 6d CFT dual to this geometry. A second possibility is to complete our solutions within globally well-defined $\mathrm{AdS}_3$ geometries. This is currently work in progress \cite{FLP}.

\section{Brief summary of 2d quiver CFTs}
\label{summary2dCFT}

The 2d (0,4) CFTs encountered in the main body of the paper are of the type recently studied in \cite{Lozano:2019jza,Lozano:2019zvg}, associated to D2-D4-NS5-D6-D8 brane set-ups. 
 These CFTs are described by (0,4) superconformal quivers with gauge groups associated to stacks of D2 and D6 branes (the latter wrapped on 4d manifolds) stretched between NS5-branes. The quivers are {\it planar}, in the sense that they consist on two long linear quivers, built out of the gauge groups associated to the D2 and D6 branes, coupled by matter multiplets. Each linear quiver consists on (4,4) gauge groups connected horizontally by (4,4) bifundamental hypermultiplets. They couple to each other through (0,4) hypermultiplets (vertically) and (0,2) Fermi multiplets (in the diagonals). These render the final {\it planar} quiver  CFTs (0,4) supersymmetric. 
Since the 2d theory is chiral one needs to be careful with gauge anomaly cancellation. This is ensured adding adequate flavour groups at each node, coming from D4 and D8-branes, that couple through (0,2) Fermi multiplets with the corresponding gauge nodes.
The gauge anomaly cancellation conditions for the $SU(N_2^{(j)})$ and $SU(N_6^{(j)})$ gauge groups are (the reader is referred to \cite{Lozano:2019zvg} for more details),
\begin{equation}
\label{anomaly}
2N_2^{(j)}=N_2^{(j-1)}+N_2^{(j+1)}+\Delta N_4^{(j)}\, , \qquad 2N_6^{(j)}=N_6^{(j-1)}+N_6^{(j+1)}+\Delta N_8^{(j)}.
\end{equation}

\begin{figure}[http!]
\centering
\includegraphics[scale=0.7]{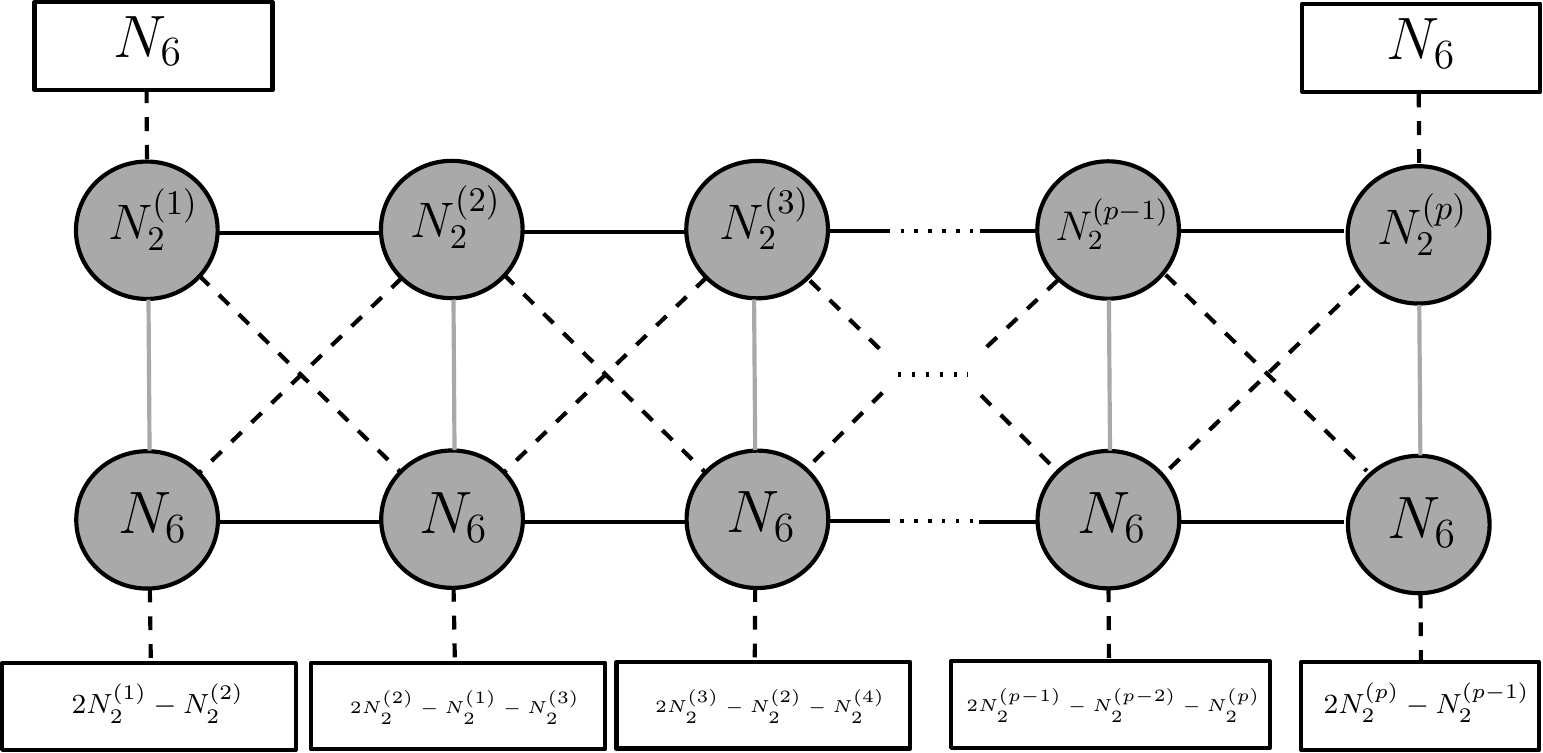}
\caption{Generic quiver field theory associated to D2-D4-NS5-D6 brane set-ups.}
\label{quiverI}
\end{figure}

In the particular situation discussed in this paper we are concerned with massless IIA supergravity. In this case there are no D8-branes in the constructions in \cite{Lozano:2019zvg} and the number of D6-branes remains constant. Therefore one of the rows of the quivers contains gauge groups of constant ranks. The general structure of these quivers is depicted in Figure \ref{quiverI}. These quivers were recently discussed in \cite{Lozano:2020bxo}.

\section{Massive IIA origin of 6d Romans supergravity}\label{6dsugra}

In this appendix we discuss the consistent truncation of massive IIA string theory around the $\mrm{AdS}_6$ vacuum discovered in \cite{Brandhuber:1999np}, describing the near-horizon limit of the D4-D8 brane system. This truncation produces  in its minimal realisation, i.e. when only the supergravity multiplet is retained, a 6d gauged supergravity preserving $\ma N=(1,1)$ supersymmetry. This theory is usually called Romans supergravity. Its field content is given by the 6d gravitational field, a real scalar $X_6$, a 2-form gauge potential $\ma B_{(2)}$, three $\mrm{SU(2)}$ vectors $\ma A_6^i$ and one Abelian vector $\ma A_6^0$ \cite{Romans:1985tw}. Here we will restrict to the case of vanishing vector fields. 

The truncation from massive IIA supergravity to Romans supergravity was worked out in \cite{Cvetic:1999un}. The ansatz for the metric is characterised by an internal manifold locally realised as a fibration of a 3-sphere over a line,
\begin{equation}
 \begin{split}
  &ds^2_{10}=s^{-1/3}\,X_6^{-1/2}\,\Sigma_6^{1/2}\,\left[ds^2_6+2g^{-2}\,X_6^{2}\,ds_{4}^2\right]\ ,\\
  &ds_{4}^2=d\xi^2+\Sigma_6^{-1}\,X_6^{-3}\,c^{2}\,ds^2_{S^3}\ ,
  \label{truncationansatz6d}
 \end{split}
\end{equation}
where $\Sigma_6=X_6\,c^2+X_6^{-3}\,s^2$ with $c=\cos\xi$ and $s=\sin \xi$. The 10d fluxes are decomposed as  \cite{Cvetic:1999un}
\begin{equation}
\label{10dfluxesto6d}
 \begin{split}
  F_{(4)}&=-\frac{4\sqrt 2}{3}\,g^{-3}\,s^{1/3}\,c^3\,\Sigma_6^{-2}\,U\,d\xi\,\wedge\,\text{vol}_{S^3}-8\sqrt{2}\,g^{-3}\,s^{4/3}\,c^4\,\Sigma_6^{-2}\,X_6^{-3}\,dX_6\,\wedge\,\text{vol}_{S^3}\\
  &-\sqrt2 \,g^{-1}\,s^{1/3}\,c\,X_6^4\,\star_{\,6}\ma F_{(3)}\,\wedge\,d\xi-m\,s^{4/3}\,X_6^{-2}\,\star_{\,6}\ma B_{(2)} \,, \\
  F_{(2)}&=m\,s^{2/3}\,\ma B_{(2)}\ ,\qquad H_{(3)}=s^{2/3}\,\ma F_{(3)}+\sqrt{2}\,g^{-1}\,m\,s^{-1/3}\,c\,\ma B_{(2)}\,\wedge\,d\xi\ ,\\
  e^{\Phi}&=s^{-5/6}\,\Sigma_6^{1/4}\,X_6^{-5/4}\ ,\qquad F_{(0)}=m\,,
 \end{split}
\end{equation}
where $U=X_6^{-6}\,s^2-3X_6^2\,c^2+4\,X_6^{-2}\,c^2-6\,X_6^{-2}$ and $\ma F_{(3)}=d \ma B_{(2)}$. The 6d theory resulting from this truncation preserves 16 real supercharges and has $\mathbb{R}^+\times \mathrm{SO}(4)$ global isometry group. The two parameters $g$ and $m$ are associated, respectively, to the gauging of the $\mathrm{SU}(2)_R$ R-symmetry group, realised as the diagonal $\mathrm{SU}{(2)}$ within $\mathrm{SO}(4)$, and to a mass deformation of the 2-form. In particular, the truncation ansatz \eqref{truncationansatz6d} produces a scalar potential in six dimensions defined by the superpotential
\begin{equation}
 f_6(m,g,X_6)=\frac{1}{8}\,\left(m\,X_6^{-3}+\sqrt2 \,g\, X_6 \right)\, ,
 \label{6dsuperpotential}
\end{equation}
where the two gauging parameters are linked as $m= \frac{\sqrt{2}}{3}g$. The 6d Lagrangian has the form
\begin{equation}
\begin{split}
\label{6dlagrangian}
 \ma L_6&=  R_{\,6}-4 \, X_6^{-2}\,\star_{\,6}\,dX_6\,\wedge \, dX_6-\frac12\,X_6^4\,\star_{\,6}\,\ma F_{(3)}\wedge \ma F_{(3)}-V_6\\
 &- m^2\, X_6^{-2} \, \star_{\,6}\,\ma{B}_{(2)}\wedge \ma{B}_{(2)}\,-\frac13\,m^2 \, \ma{B}_{(2)}\,\wedge \,\ma{B}_{(2)}\,\wedge \ma{B}_{(2)}\,,
 \end{split}
\end{equation}
where the scalar potential is given by
\begin{equation}
 V_6 = 16\,X_6^{2}\,\left(D_{X}f_6\right)^{2} -80\,f_6^{2}\,.
\end{equation}
The 6d theory \eqref{6dlagrangian} has a supersymmetric $\mathrm{AdS_6}$ vacuum at $X_6=1$ and vanishing gauge potentials. This vacuum corresponds to the string vacuum of the D4-D8 set-up introduced in \cite{Brandhuber:1999np}. In this case the 4d internal manifold in \eqref{truncationansatz6d} becomes a round 4-sphere\footnote{More precisely this is the upper hemisphere of a 4-sphere with boundary at $\xi\rightarrow 0$ \cite{Brandhuber:1999np, Cvetic:1999un}.}, and the only non-zero terms in the fluxes and dilaton given by \eqref{10dfluxesto6d} are
\begin{equation}\label{BOvacuum}
 F_{(4)}=\frac{20\sqrt{ 2}}{3}\,g^{-3}\,s^{1/3}\,c^3\,d\xi\,\wedge\,\text{vol}_{S^3}\ ,\qquad e^{\Phi}=s^{-5/6}\,.
\end{equation}
These are exactly the fluxes and the dilaton describing the near-horizon limit of the D4-D8 system introduced in \cite{Brandhuber:1999np}.

\end{document}